\documentclass{aa}
\bibpunct{(}{)}{;}{a}{}{,}
\usepackage{graphicx}
\usepackage{natbib}
\usepackage{txfonts}
\usepackage{hyperref}

\begin{document}

\title{Auto-correlation functions of astrophysical processes, and their relation to Gaussian processes}

\subtitle{Application to radial velocities of different starspot configurations}

\author{M.~Perger\inst{1,2}
       \and G.~Anglada-Escud\'e\inst{1,2}
        \and I.~Ribas\inst{1,2}
        \and A.~Rosich\inst{1,2}
        \and E.~Herrero\inst{1,2}        
        \and J.~C.~Morales\inst{1,2}
}

\offprints{M.~Perger, \email{perger@ice.cat}}
\institute{
        \inst{1}Institut de Ci\`encies de l'Espai (ICE, CSIC), Campus UAB, Carrer de Can Magrans s/n, 08193 Bellaterra, Spain\\
        \inst{2}Institut d'Estudis Espacials de Catalunya (IEEC), 08034 Barcelona, Spain
}

\date{Received: \, \today \, Accepted: ...}

\abstract
{Accounting for the effects of stellar magnetic phenomena is indispensable to fully exploit radial velocities (RVs) obtained using modern exoplanet-hunting spectrometers. Correlated time variations are often mitigated by non-trivial noise models in the framework of Gaussian processes (GPs). These models rely on fitting kernel functions that are motivated on mathematical grounds, and whose physical interpretation is often elusive.}
{We aim to establish a clear connection between stellar magnetic activity affecting RVs and their corresponding correlations with physical parameters, and compare this connection with kernels used in the literature.}
{We use simple activity models to investigate the relationship between the physical processes generating the signals and the covariances typically found in data, and to demonstrate the qualitative behaviour of this relationship. We use the {\tt StarSim} code to calculate RVs of an M dwarf with different realistic evolving spot configurations. The auto-correlation function (ACF) of a synthetic data set shows a very specific behaviour and is explicitly related to the kernel. Gaussian process regression is performed using the quasi-periodic (QP) and simple harmonic oscillator (SHO) kernels of the {\tt george} and {\tt celerite} codes, respectively. Comparison of the resulting kernels with the exact ACFs allows us to cross-match the kernel hyper-parameters with the introduced physical values, study the overall capabilities of the kernels, and improve their definition.}
{We find that the QP kernel provides a more straightforward interpretation of the physics. It is able to consistently recover both the introduced rotation period $P_{\rm rot}$ and the spot lifetime. Our study indicates that the performance can be enhanced by fixing the form factor $w$ and adding a physically motivated cosine term with period $P_{\rm rot}/2$, where the contribution to the ACF for the different spot configurations differs significantly. The newly proposed quasi-periodic with cosine (QPC) kernel leads to significantly better model likelihoods, can potentially distinguish between different spot configurations, and can thereby improve the sensitivity of RV exoplanet searches.}
{}


\maketitle

\section{Introduction} \index{int} \label{int}

High-resolution Doppler spectroscopy has been instrumental in finding and confirming the existence of planetary companions orbiting stars. Modern instruments like HARPS \citep{2003Msngr.114...20M}, CARMENES \citep{2018SPIE10702E..0WQ}, and ESPRESSO \citep{2010SPIE.7735E..0FP} attain radial velocity (RV) precisions in the m\,s$^{-1}$ domain and better. This is effectively comparable to astrophysical variations even in the case of stars with moderate to low magnetic activity levels. Therefore, a precise understanding of intrinsic stellar phenomena is mandatory in order to expand the search for low-mass, rocky exoplanets.

In the cm\,s$^{-1}$ domain, stellar effects include short-term variability induced by surface granulation and stellar oscillations \citep[see e.g.][]{2011A&A...525A.140D,2019ApJ...874..107M}. Furthermore, magnetic phenomena produce cool spots and hot faculae co-rotating with the stellar surface and with lifetimes of the order of a fraction to several rotational periods. These phenomena introduce RV signals that can distort, mimic, and even hide the Keplerian signature of a planetary companion. The filling factor (i.e. the fraction of surface covered with spots) and the location of the spots can also change over time producing long-term magnetic cycles of the order of years to decades \citep{2018A&A...612A..89S}. Magnetic effects are found to be even larger for cooler stars (M type), which are the prime targets for several RV surveys conducted over the last decade \citep[e.g.][]{2004A&A...423..385P, 2016A&A...593A.117A, 2018A&A...612A..49R} as they offer a shortcut to finding temperate rocky planets. An overview of the impact of various stellar effects on RV measurements is provided in \citet{2012Natur.491..207D}.

The scientific community has put a lot of effort into the treatment of star-induced RV variability. Since its first use by \citet{2014MNRAS.443.2517H}, and in particular when the first results of the RV challenge by \citet{2017A&A...598A.133D} were published, Gaussian process (GP) regression \citep{2015MNRAS.452.2269R, Roberts20110550} has become one of the most commonly used tools  to model and mitigate correlated effects in RV time series arising from stellar activity. The flexibility of GP algorithms makes them suitable for accounting for the effects of rotating spots on the data using a kernel function defined by a few hyper-parameters and the application of optimisation techniques (e.g. maximisation of the likelihood function). However, the association of the mathematical kernel hyper-parameters with true physical phenomena has not been clearly discussed and evaluated.

We have developed the {\tt StarSim} code \citep{2016A&A...586A.131H, Rosich_2020} to investigate the impact of stellar activity on time-series data using a physical approach. In its forward model functionality, {\tt StarSim} assumes a stellar surface configuration, which is mapped onto a grid of elements either representing the immaculate photosphere, a cool dark spot, or a hot facula surrounding the spot. Observables (photometric and spectroscopic) at a certain epoch are predicted by integrating the contribution of all elements over the full visible stellar surface, including geometric effects, convective shifts, line asymmetries, characteristics of the spots and faculae, and of the host star, including the possibility of differential rotation.

In this study, we create synthetic RV time-series data of a rotating spotted star using different realistic, evolving spot distributions and the {\tt StarSim} code. To interpret the simulated data, we analyse the auto-correlation function (ACF) and fit different commonly used GP kernels. We are then able to study the connection of the introduced physical parameters with the posterior distribution of the GP hyper-parameters. We present the mathematical background to the likelihood function and the ACF in Sect.\,\ref{s2}. In Sect.\,\ref{s1}, we introduce the {\tt StarSim} RVs of the different spot distributions and analyse their ACFs, and in Sect.\,\ref{s3}, we apply GP regression using the different kernels and evaluate their performance. The results are summarised in Sect.\,\ref{s4}.

\section{Auto-correlation function and kernels}  \index{s2} \label{s2}

\subsection{Likelihood function}  \index{s2a} \label{s2a}

The goal of a regression exercise is to include a parameterised representation of the data (model) that best fits the variability measured in the observations. In the current literature, this is done by maximising a merit function called the likelihood function $\mathcal L$. The likelihood function represents the probability distribution of the data (a time series $v$ and a function of time $t$ in our case) to occur given a specific model, which can contain a number of adjustable parameters. The likelihood function arises from the common assumption that the \textit{residuals} $r_i$, that is, the difference between the observation $v_{\rm obs}(t_i)$ and a parameterised model $v_{\rm mod}(t_i)$, are drawn from a joint multivariate Gaussian distribution and is usually defined as 
\begin{eqnarray}
\mathcal L&=&\frac{1}{(2\pi)^{N/2}} |C|^{-1/2} \rm exp \Big( -\frac{1}{2} \sum_{i=1}^{N} \sum_{j=1}^{N} r_{i} r_{j} C_{ij}^{-1} \Big), \\ \index{X1}  \label{X1}
r_i&=&v_{\rm obs}(t_i)-v_{\rm mod}(t_i), \index{X2}  \label{X2}
\end{eqnarray}
\noindent 
where $N$ is the number of observations, and $C$ is the covariance matrix between data residuals. This covariance matrix can be written as
\begin{eqnarray}
C_{ij} &=& (\sigma^{2} + \epsilon_{i}^2 ) \, \delta_{ij} + K(t_i-t_j) \index{X4}  \label{X4}
,\end{eqnarray}
\noindent where $\epsilon_i$ is the uncertainty on the measurement $v(t_i)$, $\sigma$ represents uncorrelated random noise (jitter), and $\delta_{ij}$ the Kronecker Delta, and covariances between data points are specified by the so-called stationary kernel function $K$ (or simply \textit{kernel} hereafter); for a detailed discussion see for example \citet{2013MNRAS.429.2052B, 2013A&A...556A.126A, 2018Natur.563..365R}. For numerical reasons, it is more practical to work with the logarithm of the likelihood function $\ln{\mathcal L}$, which will also be maximal for the same parameter values as the likelihood function. The maximum value of a likelihood function that depends on a number of adjustable parameters $n_p$ will always improve when adding more. When comparing models with different numbers of parameters, the Bayesian information criteria (BIC) is often used as a way to apply a penalty to more complex models, and can be computed as
\begin{equation}
{\rm BIC} = n_p\, \ln{N} + 2\,\ln{\mathcal L}. \index{X93}  \label{X93}
\end{equation}
\noindent Although this indicator does not account for Bayesian priors nor multi-variate local minima, it is used below to additionally evaluate the relation between different models with increasing complexity.

\subsection{Relation between physical models and correlated noise}

It is important to note that, in this picture, our interpretation of the data can be included in two different parts of the likelihood function. Firstly, adjustable parameters can be in the directly parameterisable model $v_{\rm mod}$, and secondly in the covariance matrix $C$ and its kernel $K$. Whether a signal is better described as part of the $v_{\rm mod}$-model (e.g. detection of a planetary signal) or inside the covariance matrix $C$ is often used as a diagnostic to decide if the signal is induced by a real planet or is caused by stellar activity. It has been acknowledged that there must be a certain amount of degeneracy between signals described by $v_{\rm mod}$ and those in the covariances, which is the source of all sorts of controversial claims. In this section, we re-derive a basic description of correlations between data sets and show the relationship between these correlations and the signals that generate them, which should provide us with useful information for subsequent discussions.

The covariance between the values of $v=v(t)$ and the same variable at a later time $v^\prime = v(t^\prime)$ is defined as
\begin{eqnarray}
C \left[v,v^\prime\right]  &=& 
E\left[ 
   \left(
      v - E[v]
   \right) 
   \cdot 
   \left(
      v^\prime-E[v^\prime]
   \right) 
\right] 
\label{eq:K}
,\end{eqnarray}
\noindent where the expected value $E$ of a variable $v$ is defined as
\begin{eqnarray}
E[v] = Z^{-1} \int v(t)\, dt
,\end{eqnarray}
\noindent and $Z = \int dt$ normalises the integral for a relevant time interval. Formally, this integral should expand from $-\infty$ to $\infty$, but it can be defined over a finite interval in which case $Z=T$, where $T$ is the time-span of the observations, as shown in subsequent examples. Assuming that the function $v(t)$ is non-zero over a finite interval, the expected values satisfy $E[v] = E[v^\prime] = \gamma$, which is a constant (average of the time-series). By explicitly computing the products before the expected values, substituting $\gamma$ into Eq.\,(\ref{eq:K}), and defining $K(t-t^\prime) = C[v,v^\prime]$ we obtain
\begin{eqnarray}
K(\tau) &=& E[v \cdot v^\prime] - \gamma^2  \label{eq:Kernel} ,\\
\tau &=& t - t^\prime
,\end{eqnarray}
\noindent which represents the correlations induced by the physical processes described by $v(t)$ as a function of the time-lag $\tau$. Therefore, this is the same object as the kernel function in Eq.~\ref{eq:K}. From this expression we can immediately note that a non-zero value for the expected $\gamma^2$ will contribute to the kernel unless it is fitted as a free parameter in the explicit part of the model (e.g. a constant offset in $v(t)$). Another fact that becomes obvious when looking at this expression is that the kernel is directly connected to the physical process $v(t)$ through the auto-correlation function of the generating signal, which can be explicitly computed using
\begin{eqnarray}
K(\tau) &=& \frac{1}{Z} \int v(t) \, v(t-\tau) \,dt \label{eq:acf}
,\end{eqnarray}
\noindent and can be computed for any generating function $v$ given from models, or can be approximately computed from observations themselves (the integral becomes a discrete sum). This feature allows the development of physically motivated kernels by comparing simulated observations (signals generated by spots on rotating stars) to the kernels typically used in the literature.

Before investigating realistic cases, we illustrate the process of how to compute exact kernels in two simple but representative situations. The two generating functions discussed here -- a linear trend and a sinusoidal signal -- may not necessarily be used in an optimisation algorithm, as they can be more easily added in the explicit model, but we think they provide useful insights into the connection between physics and kernels, and provide clues as to the sources of degeneracy often encountered in the literature.
\newline 

\noindent \textbf{Case 1 -- A linear trend}. In this case, the generating function corresponds to $v(t)=a + b t$, where $a$ is a constant offset and $b$ is the slope. Substituting in Eq.~\ref{eq:acf}, the corresponding kernel becomes
\begin{eqnarray}
K_{\rm lin}(\tau) 
&=&  \frac{1}{Z}\left[ 
\int_0^T 
\left(a + b t \right)
\left(a + b (t - \tau) \right) dt 
\right] \nonumber
\\ 
&=& 
a^2 + 
a b \,T +
\frac{b^2} {3}\,T^2 -
\big( a b + \frac{b^2 T }{2} \big)\,\tau \label{eq:onesidedlin}
,\end{eqnarray}
\noindent which only contains constants and linear terms with $\tau$. Therefore, a suitable parameterised kernel could simply be expressed as  
\begin{eqnarray}
K_{lin}(\tau) = \alpha + \beta \tau
,\end{eqnarray}
\noindent where $\alpha$ and $\beta$ would be related to the slope and offset of the time-series in a complicated way following Eq.\,(\ref{eq:onesidedlin}).
\newline

\noindent \textbf{Case 2 -- A sinusoidal signal}.
The generating function has now the form $v(t)= \kappa \sin w t$. Again substituting in Eq.~\ref{eq:acf}, the resulting kernel is
\begin{eqnarray}
K_{\rm sin}(\tau) &=&
\frac{1}{Z} \int
\kappa^2 
\sin \left(\omega t\right) \,
\sin \left(\omega (t - \tau ) \right)
\, dt \nonumber \\ 
& = & \kappa^2 \cos (\omega \tau)  + \, O[\kappa^2/N]  \label{eq:cosK}
,\end{eqnarray}

\noindent where $\kappa^2$ (square of the semi-amplitude) and $\omega$ (angular frequency corresponding to a period $P = 2\pi/\omega$) would be the adjustable free parameters. To derive this expression we used trigonometric identities, and assumed that the integrals of the trigonometric functions over the time-domain go towards zero when several cycles $N \sim T/P$ are covered, as reflected in the last term in Eq.\,(\ref{eq:cosK}). This latter assumption may not hold when examining signals with periods close to the baseline of the observations, but the expression still provides an approximate idea of the functional shape of the kernel. The same calculation can be done adding an arbitrary phase $\phi$ as $v(t)= A \sin (\omega t + \phi)$, reaching an identical result independently of the value of $\phi$. This leads to the very general conclusion that sinusoidal signals always produce cosine-like correlations. This example illustrates that, for a general signal, $\kappa^2$ is an approximation of the quadratic sum of all the terms in the Fourier decomposition of a signal at $\tau=0$. Therefore, it is useful to name this new quantity $\kappa_0 = \sqrt{K(\tau=0)}$, which essentially represents the combined amplitude of the signals causing correlations.
\newline

Correlations affecting precision Doppler measurements are not necessarily as simple as these cases. Moreover, observed correlations are likely to be the result of the superposition of several generating functions. One can easily prove that, while physical signals are additive, their resulting correlations do not combine linearly. As an example, if a time-series contains two signals $x(t)$ and $y(t)$, an appropriate physically motivated kernel should consist of
\begin{eqnarray}
K_{xy}(\tau) &=& 
{\rm C}[x,x^\prime] + {\rm C}[y,y^\prime] + {\rm C}[x,y^\prime] + {\rm C}[y,x^\prime], 
\end{eqnarray}
\noindent which contains the ACFs of both signals (first two terms on the right-hand-side), plus the covariances of the cross-terms (two last terms), thus resembling the effect of physical interference. The same principle generalises to more than two signals. That is, the total covariance of a generating signal always contains the auto-correlation terms first (which quadratically contribute to $\kappa_0$), and all pairs of cross terms which tend to chaotically cancel out at large time-lags $\tau$ if they are not strictly periodic. 

\section{Synthetic radial velocities of an M dwarf as case study}  \index{s1} \label{s1}

To generate test data, and as such stars are still the prime targets in the search for rocky exoplanets, we adopt an M-dwarf star with $T_{\rm eff}=3750$\,K, $\log{\rm g}=4.5$, and [Fe/H]=0.0 rotating non-differentially with a rotation period of $P_{\rm rot}=25$\,days. We assume dark spots to have a temperature 300\,K cooler than the photosphere, roughly following  \citet{2018A&A...614A..35M}, and we consider no faculae. Spots in the {\tt StarSim} code are defined by five characteristic parameters, namely longitude, colatitude, radius, time of appearance, and lifetime. We first run a suite of experiments to obtain a basic understanding of the ACFs induced by single spots (Sect.~\ref{sec:singlespot}), and then we proceed with multi-spot configurations (Sect.~\ref{sec:multispot}). Commonly used kernels are fitted to the simulated data and discussed in Sect.~\ref{s3}.

\subsection{Single-spot configurations}   \label{sec:singlespot}

\begin{figure}
        \centering
        \includegraphics[width=\linewidth]{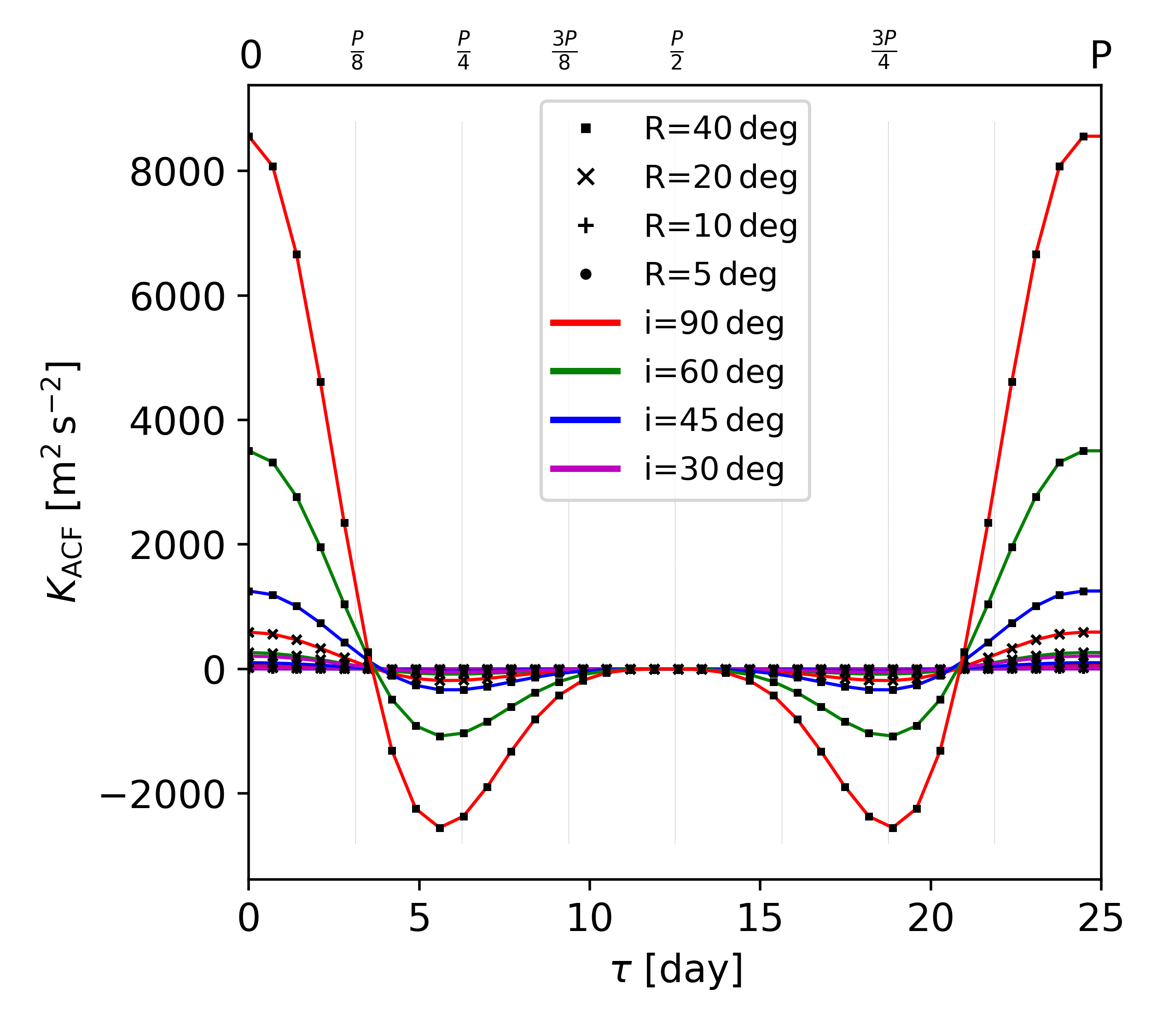}
        \caption{Auto-correlation functions of 144 RV data points of a rotating star ($P_{\rm rot}=25$\,days) with one single central spot. We show the dependence of $K_{\rm ACF}$ on stellar inclination angles $i=$ 30, 45, 60, and 90\,deg with the colours as indicated. The dependence on the radius of the dark spot $R=$ 5, 10, 20, and 40\,deg is illustrated by the different symbols as indicated. We show the repetitive part of the ACFs with $\tau < P_{\rm rot}$.}
        \label{ACF1}
\end{figure}

In a first numerical experiment we create RV time-series as induced by a single central dark spot on the stellar surface with a variety of stellar inclinations and spot radii in order to study their effect on the ACF. Radial velocity sets consist of 144 epochs over a 100-day time-span (every 1000\,min), stellar inclinations of $i$=30, 45, 60, and 90\,deg, and spot radii of $R$=5, 10, 20, and 40\,deg. We assume no noise or measurement uncertainty. In Fig.\,\ref{ACF1}, the repetitive parts ($\tau < P_{\rm rot}$) of the ACFs of these datasets are shown in different colours for the various values of $i$, and for the different $R$ with the different symbols, as labelled. The correlation decreases as both parameters decrease. In general, $K_{\rm ACF}$ is maximal at multiples of $P_{\rm rot}$ (25\, days) and close to zero at $P_{\rm rot}/2$. Furthermore, the curves are defined by two characteristic symmetric lobes of negative correlation. The time-lags for which a minimum of $K$ and $K=0$ are achieved and the depth of those lobes are defined by the description of convection and limb darkening of the {\tt StarSim} code.

\begin{figure}
        \centering
        \includegraphics[width=\linewidth]{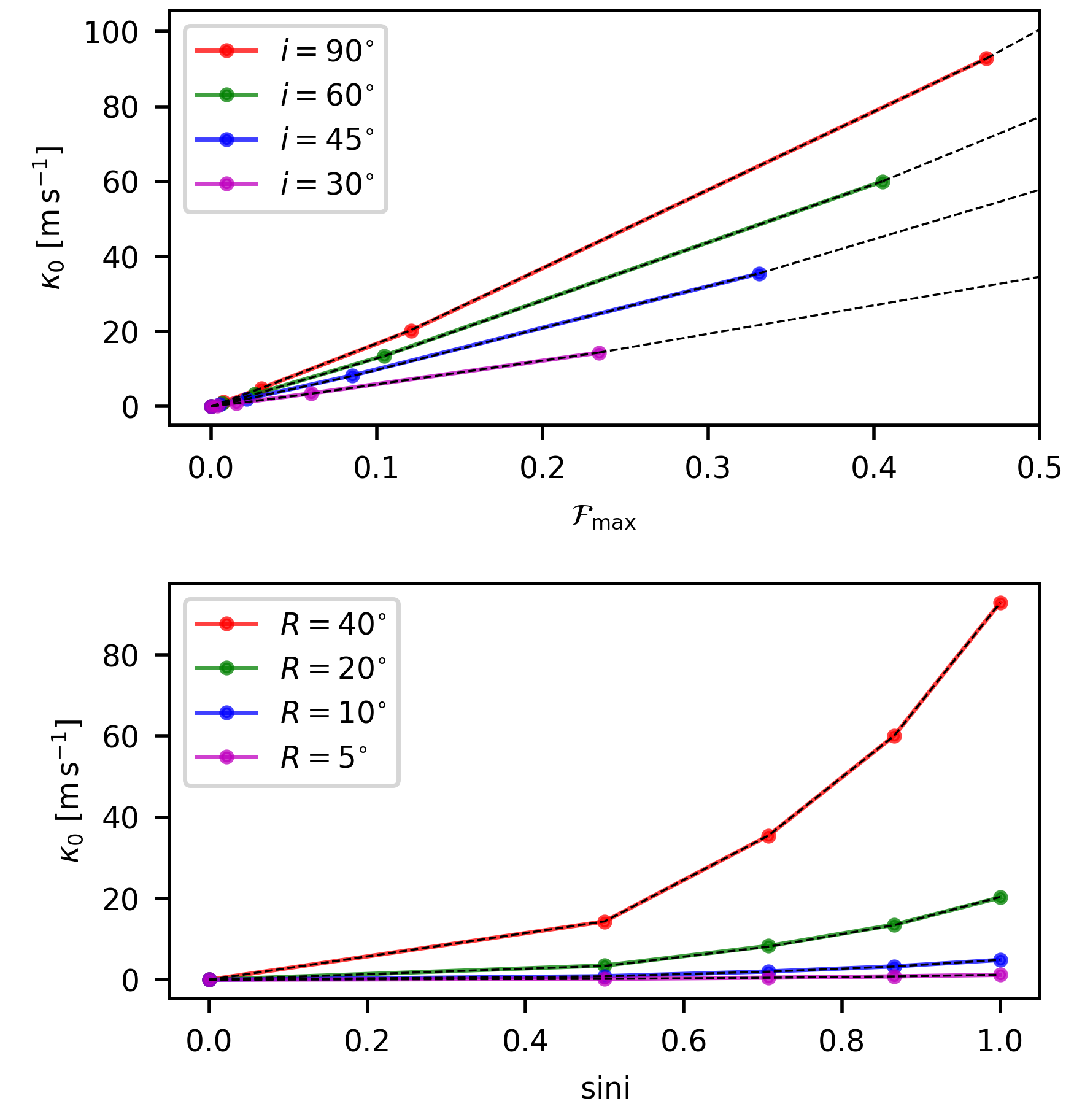}
        \caption{Square root of the maximum ACF value of our test-case M dwarf at $\tau=0$, $\kappa_0$, as a function of the maximum spot filling factor, $\mathcal F_{\rm max}$ (top panel), and of the sine of the stellar inclination angle, $\sin{i}$ (bottom panel). The curves are shown with colours as indicated for the different inclinations and different spot radii. The black dashed lines are the best fits on the curves.}
        \label{h_calib}
\end{figure}

In Fig.\,\ref{h_calib} we show $\kappa_0$, linearly correlated with the RVs of our test case, as a function of the maximum projected spot coverage $\mathcal F_{\rm max}$ (top panel) and $\sin{\rm i}$ (bottom panel) for each of the spot radii and inclination angles as indicated, and including theoretical values for $i=0$\,deg and $R=0$\,deg. The maximum filling factor of the spots $\mathcal F_{\rm max}$ is calculated for the different spot radii $R$, inclinations $i$, and colatitudes $\theta$ (=90\,deg) as
\begin{equation}
\mathcal F_{\rm max} = 2 \, (1 - \cos R) \, (\cos \theta \, \cos i + \sin \theta \, \sin i).     \index{a1}  \label{a1}
\end{equation}

The dependence of the RV amplitude on $\mathcal F_{\rm max}$ is close to linear, with only the increasing projection distortion of larger spots causing some slight curvature. However, the dependence on $\sin i$ is more parabolic. For an analytic representation of our test case, we could obtain a reasonable fit with a product of a second-order polynomial in $\mathcal F_{\rm max}$ and a fourth-order polynomial in $\sin{\rm i}$ (black dashed curves in Fig.\,\ref{h_calib}). However, we note that the details of such a fit depend on the spectral properties in a convoluted way. Our recommendation when analyzing a dataset from a star with at least one suspected spot is to perform a similar quick simulation to gauge the magnitude of $\kappa_0$ in the RV time-series. As shown below, the value of this $\kappa_0$ shall match the \textit{amplitude} parameters of all of the used kernels.

\subsection{Multi-spot configurations}  \index{sec:multispot} \label{sec:multispot}

To study situations closer to the more complex spot patterns of real stars, we calculate RV time-series of a variety of spot distribution models. The different evolving spot maps are used to create 576 synthetic RV data points with values every 3.47\,days (5000\,min) for a total time of 2000\,days and without assuming noise or measurement uncertainties. By choosing this window function, we avoid complicated sampling effects, and have a realistic time baseline for single-instrument observations, and a sufficiently large number of data points. The inclination of the test star in this case is set to 90\,deg (i.e. equator on). 

The lifetime of all spots is fixed to $T_{\rm spots}=100$\,days, because this parameter is crucial for the GP regression and we aim to calibrate it in this study. The spot radius $R$ is selected, together with $N_{\rm spot}$, to create a realistic $\sigma_{\rm RV} \sim 5$\,m\,s$^{-1}$. We also include a long-term variation, mimicking a magnetic cycle, by linearly varying the number of spots from $N$ to $N$/2 to $N$ every 1\,000\,days in intervals of 100\,days  (see Table\,\ref{Spodis}). In each interval i, each of the $N_i$ spots appears at a random day ($t_{0,i}<t_i<t_{0,i}+100$\,day) and grows with a rate of 1\,deg/day, allowing for a smooth spot evolution over the 2\,000\,days. To explore the effects produced by different configurations, we create five different maps and RV sets for each spot distribution.

\begin{figure}
\centering
        {\tt ONE} -- one active longitude\\
        \includegraphics[width=0.49\linewidth]{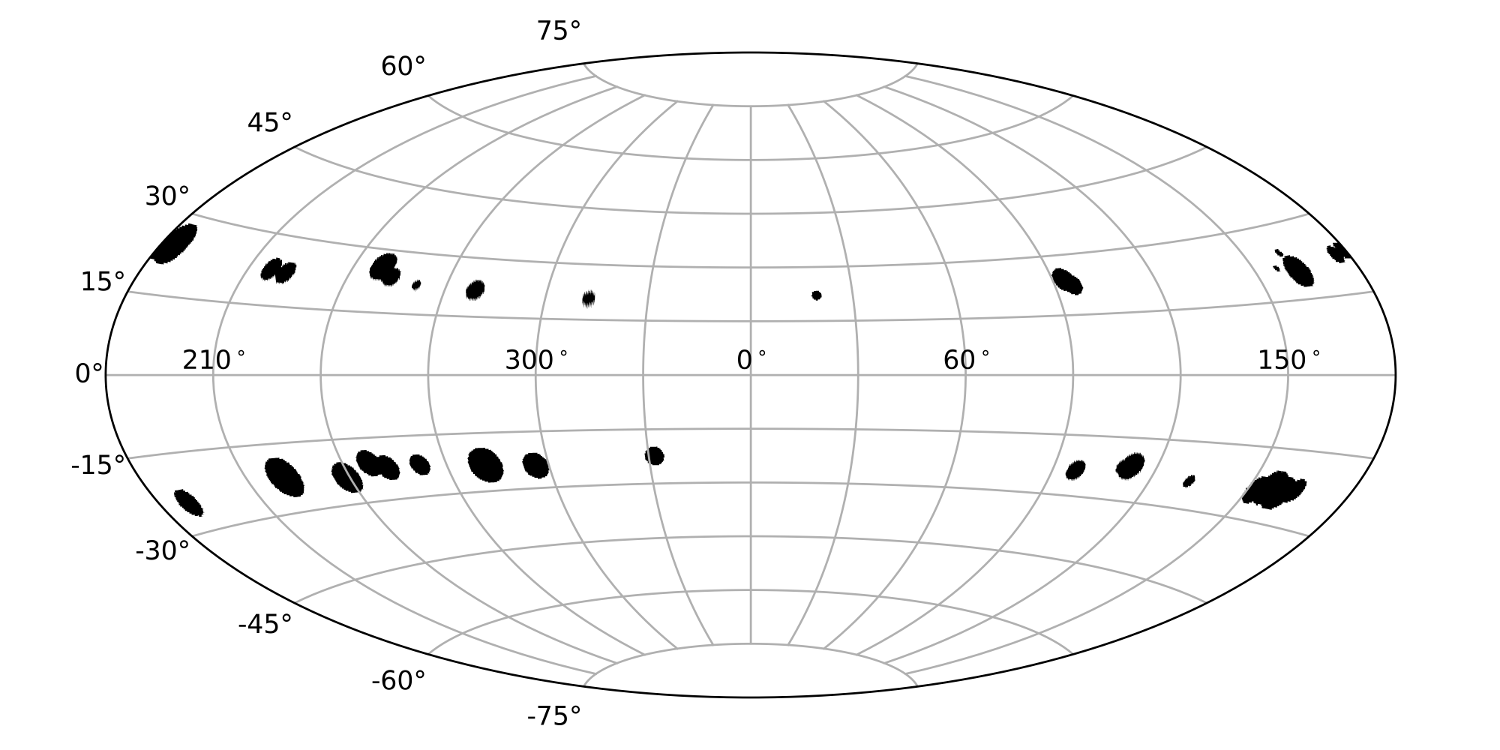}
        \includegraphics[width=0.49\linewidth]{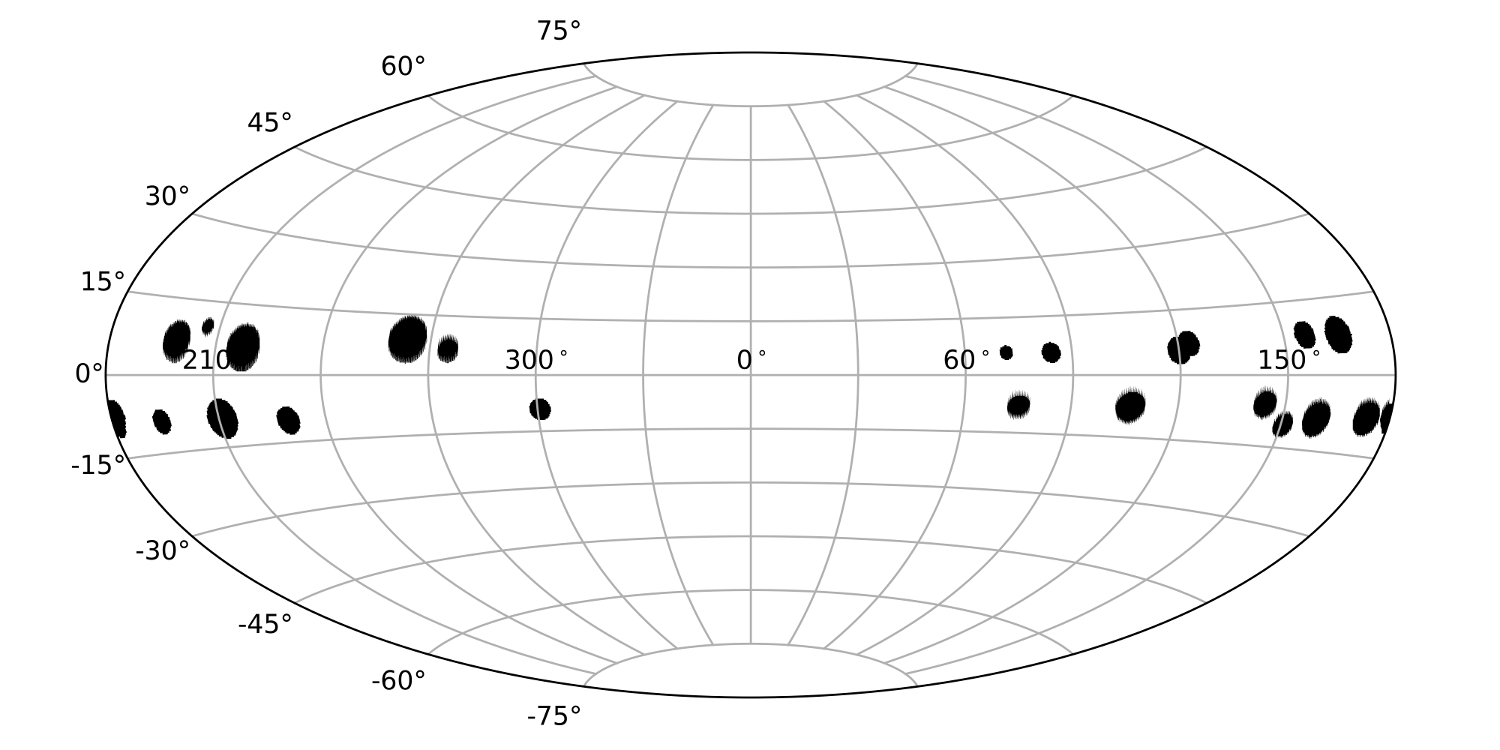}
	
        {\tt POL} -- polar spots\\
        \includegraphics[width=0.49\linewidth]{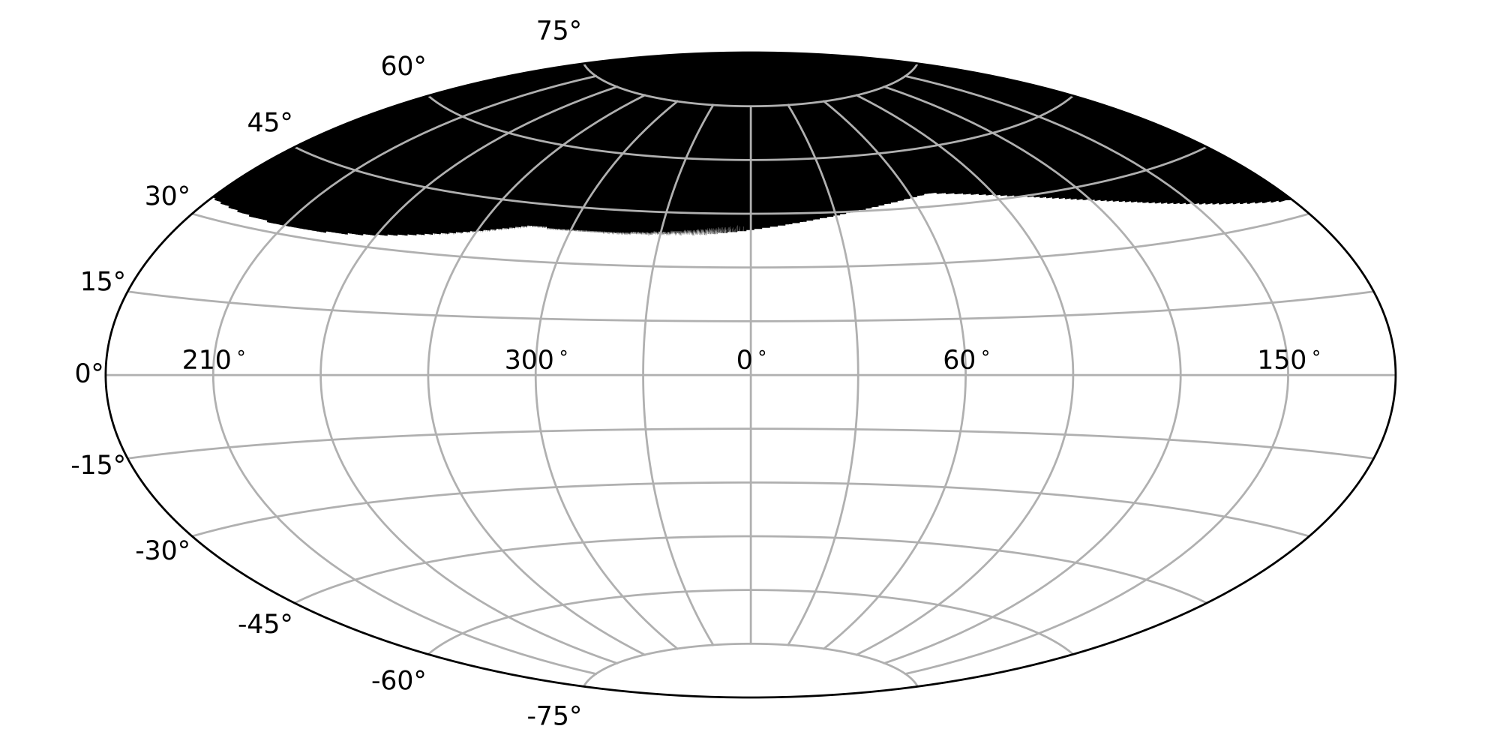}
        \includegraphics[width=0.49\linewidth]{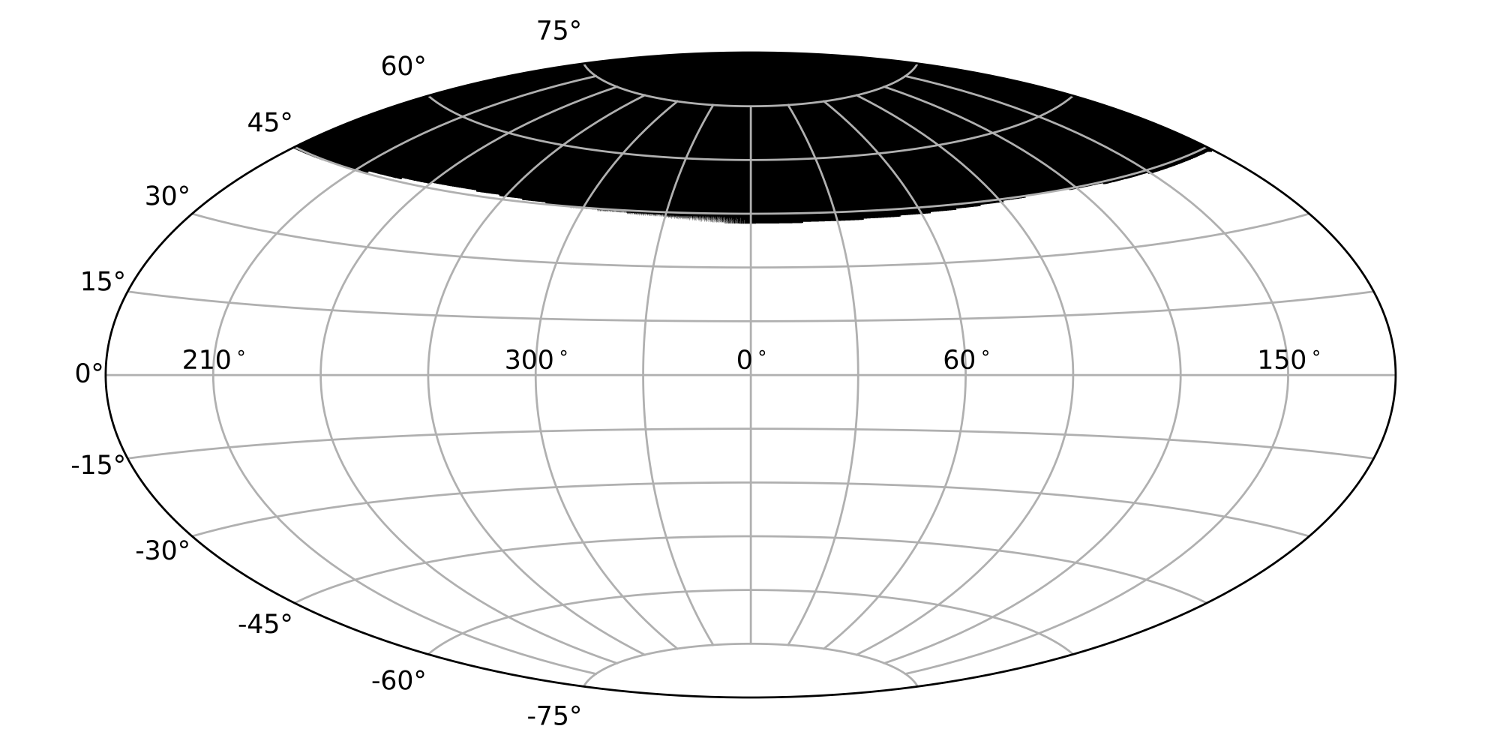}
	
        {\tt RAN} -- random distribution\\
        \includegraphics[width=0.49\linewidth]{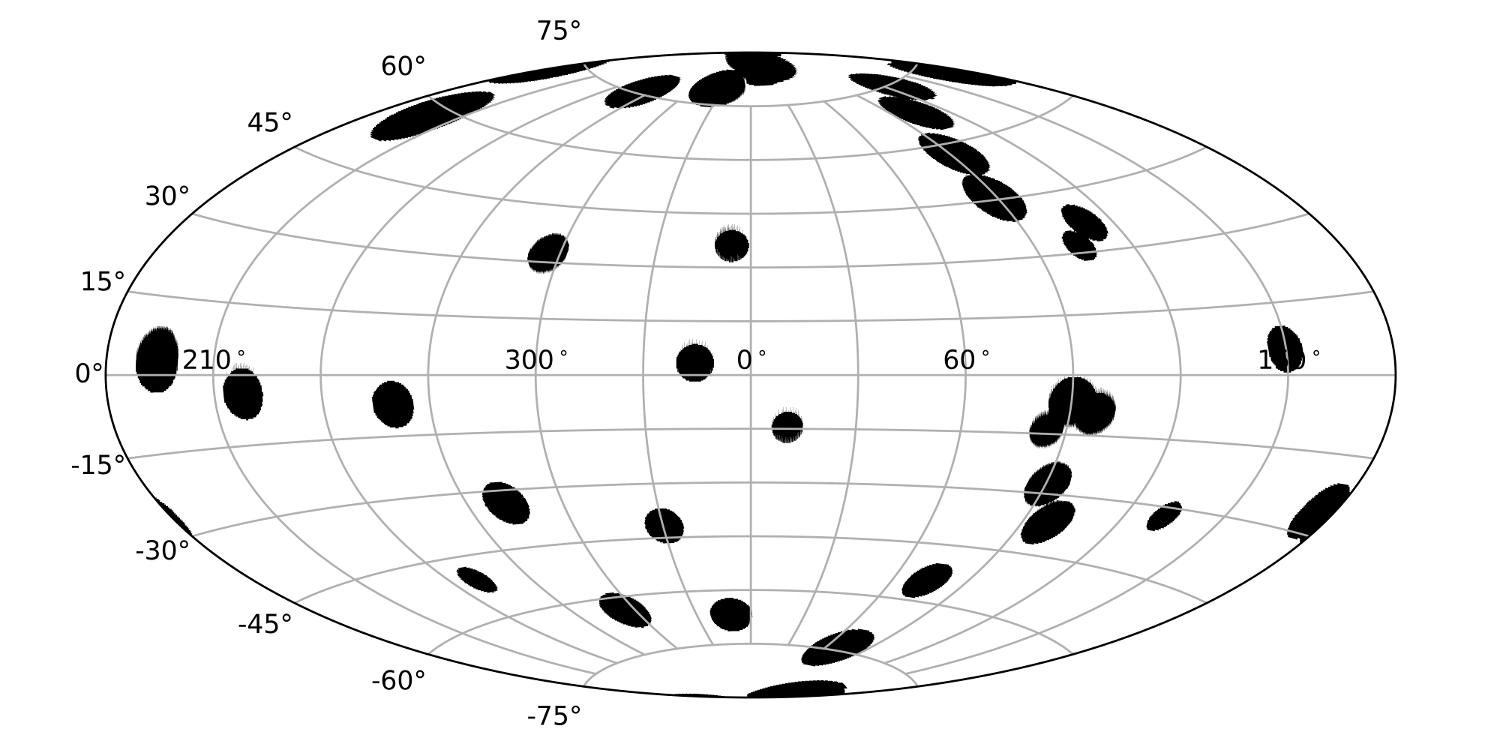}
        \includegraphics[width=0.49\linewidth]{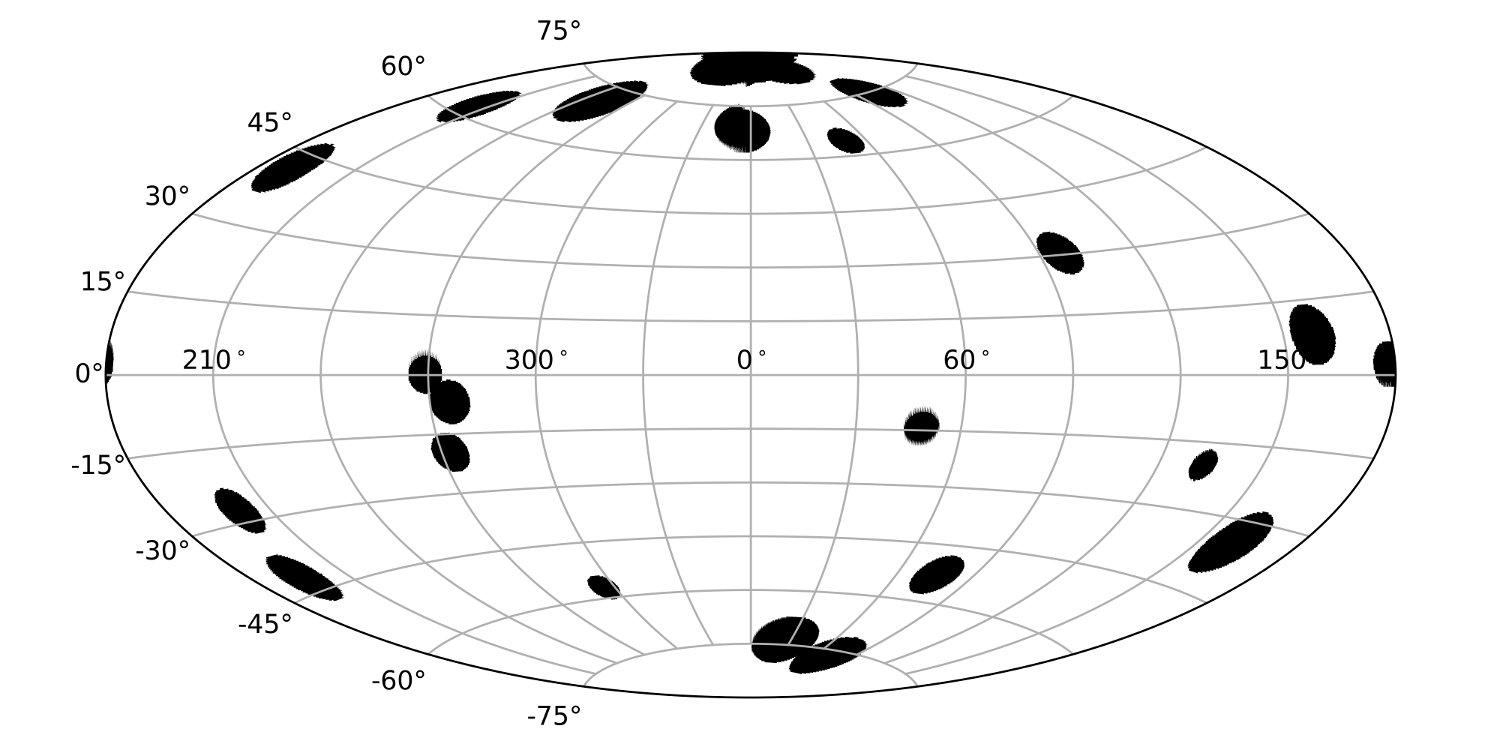}
	     
        {\tt TWO} -- two active longitudes\\
        \includegraphics[width=0.49\linewidth]{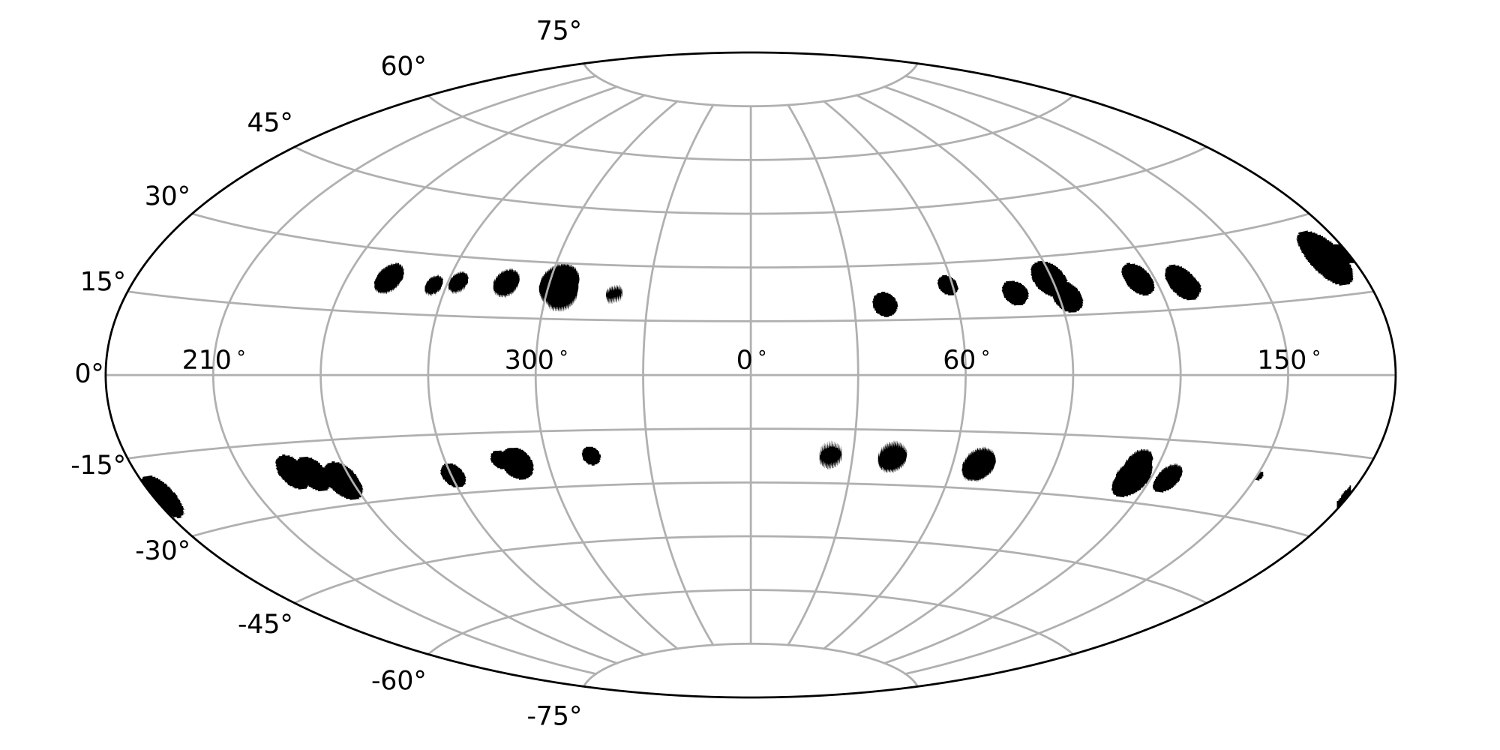}
        \includegraphics[width=0.49\linewidth]{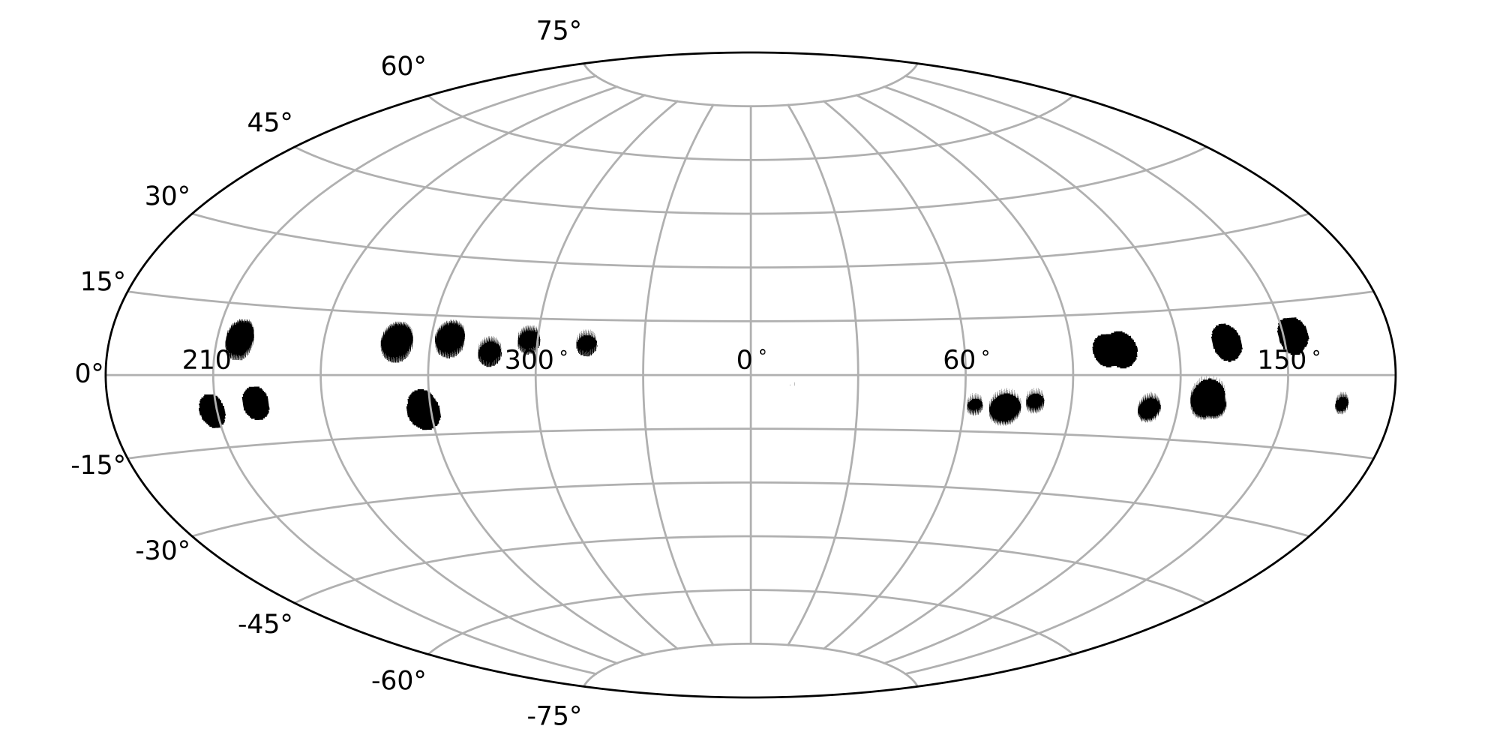}
        \caption{Spot configurations considered in this study. From top to bottom we show spot distributions with one active longitude, {\tt ONE} (active longitude at 180\,deg, $\sigma_n=60$\,deg), with axisymmetric polar spot distribution, {\tt POL}, with randomly distributed spots, {\tt RAN}, and with two active longitudes, {\tt TWO} (90 and 270\,deg, $\sigma_n=45$\,deg). The maps depicted illustrate the highest spot coverage at $t=1000$\,days (left panels) and the lowest coverage at 500\,days (right panels). Note, how the spot colatitude is changing for the {\tt ONE} and {\tt TWO} models.}
        \label{Spots1}
\end{figure}

We consider four different configurations for the distribution in longitude and colatitude of dark spots, in line with the possibilities suggested by the relevant literature on M dwarfs \citep[see e.g.][]{2008MNRAS.390..545D, 2015MNRAS.448.3053A}: 
\begin{itemize}
    \item Configuration {\tt ONE}. This configuration is inspired by the observations of the fast-rotating early-M dwarf EY\,Dra by \citet{2010AN....331..250V}, which shows a distribution of dark spots around one active longitude. To create such a spot map, we distribute a maximum of 40 small spots of 3$\pm$1\,deg radius, resulting in a total of 600 spots for a 2000-day baseline. We imitate the behaviour of solar spots, that is, the butterfly diagram, by distributing them from colatitudes of 95 to 125\,deg and from 55 to 85\,deg. Also, we place them closer to 90\,deg when the number of spots is lower. The scheme for the spot distribution is shown in Table\,\ref{Spodis}. One of the five created maps is shown in the top panels of Fig.\,\ref{Spots1} for the highest spot coverage at $t=$1000 days (left panel) and the lowest coverage at 500 days (right panel). The active longitudes of the spots are normally distributed around 180\,deg ($\sigma_{\rm n}=60\,$deg).
    \item Configuration {\tt POL.} This configuration is characterised by large spots at high latitudes (polar regions) of a star; second row of Fig.\,\ref{Spots1}). This is observed in the K-type dwarf AB\,Dor by \citet{2007MNRAS.375..567J} and using tomographic imaging of active M-type stars by \citet{2016MNRAS.461.1465H}. For this polar model we introduce only 30 spots distributed in time as shown in Table\,\ref{Spodis}. These have 45\,deg radii, and are located at colatitudes from 0 to 20\,deg, and longitudes normally distributed around 180\,deg ($\sigma_{\rm n}=90\,$deg).
    \item Configuration {\tt RAN}. A third configuration assumes a random distribution of spots characteristic of very active, fast-rotating, and fully convective stars\citep[$>$M4,][]{2011MNRAS.412.1599B}. In such cases, spots are found to be rather homogeneously distributed over the full stellar surface with high filling factors. This was shown to be the case through Zeeman Doppler imaging by \citet{2008MNRAS.390..567M, 2010MNRAS.407.2269M}. For this random distribution of spots, shown in the third row of Fig.\,\ref{Spots1}, we use 600 spots with the same magnetic cycle as for {\tt ONE}, but distribute each spot randomly over the surface. To reach the described $\sigma_{\rm RV}$=5\,m\,s$^{-1}$, we increase the sizes of spots to $5\pm1$\,deg.
    \item Configuration {\tt TWO}. In this case, we use the solar example \citep{2011A&A...529A..23Z} of two active longitudes on opposite stellar sides, which may also be the most common distribution in G-, K-, and early M-type stars \citep{2005A&A...432..657J, 2009A&A...493..193L, 2014ARep...58..478S}, and is also found for the fully convective late-M dwarf LHS\,6351 \citep{2012ARep...56..116S}. To create the spot maps, we use the same time and colatitude distribution of the 600 spots as described for {\tt ONE}, but place them on two opposite longitudes (normally distributed around 90 and 270\,deg, $\sigma_{rm n}=45$\,deg). Two representative maps (low and high spot coverage) are shown in the bottom panels of Fig.\,\ref{Spots1}.
\end{itemize}

\begin{figure}
        \centering
        \includegraphics[width=\linewidth]{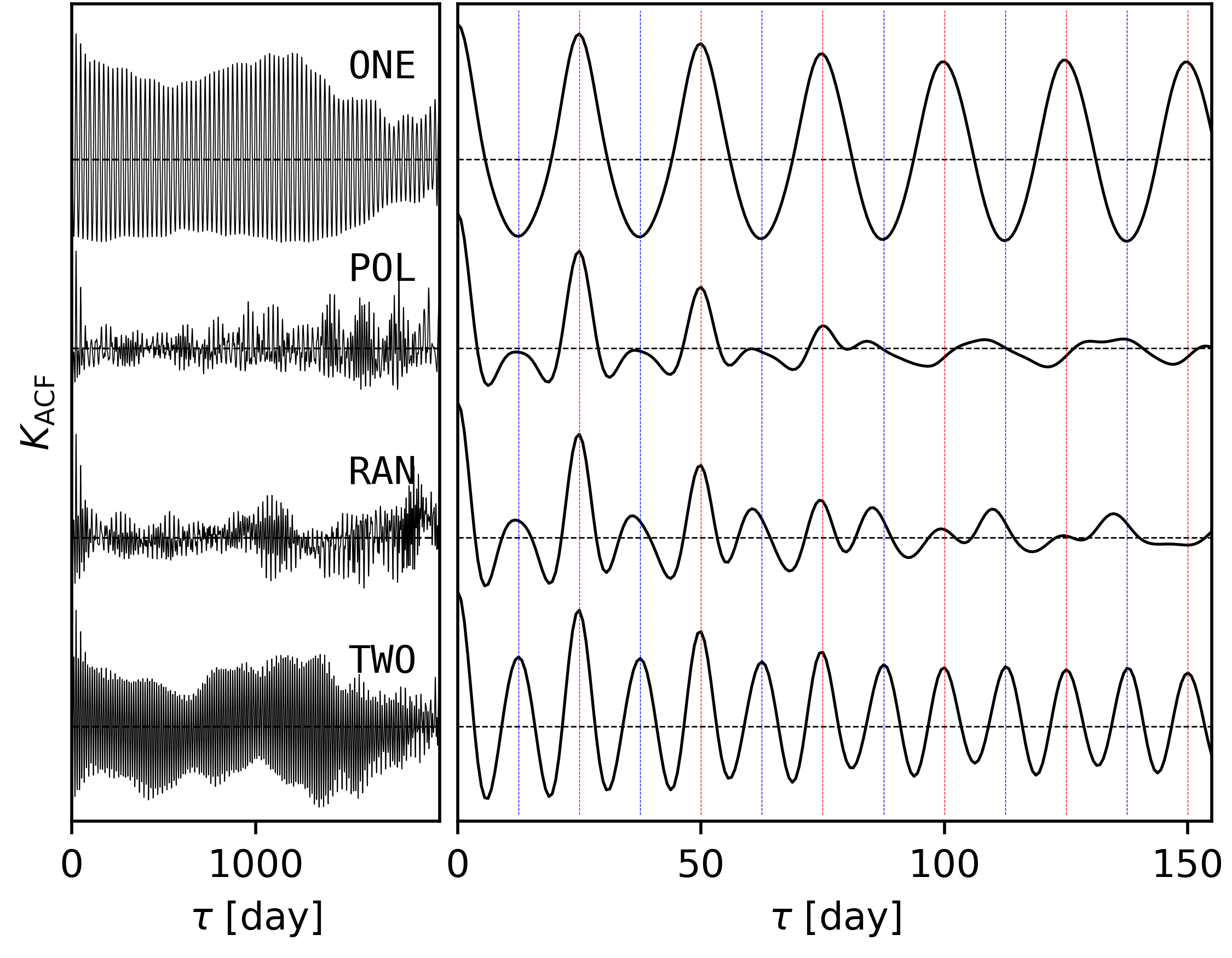}
        \caption{Auto-correlation functions of one example {\tt StarSim} RV data set for each of the four different spot distributions. Those are, from top to bottom, the {\tt ONE}, {\tt POL}, {\tt RAN}, and {\tt TWO} configurations. On the left-hand side the ACFs are shown for $0<t<2000$\,days and on the right panel for $0<t<150$\,days to highlight the injected stellar rotation at $P_{\rm rot}=25\,d$, and the spot lifetime of $100\,d$. Vertical red and blue dashed line mark $P_{\rm rot}$ and $P_{\rm rot}/2$ intervals, respectively.}
        \label{ACF2}
\end{figure}

By sampling the simulated generating function at high cadence (1\,000\,min$\sim$0.7\,d) and using numerical integration, example ACFs of all four configurations are computed, and illustrated in Fig.~\ref{ACF2}. In the left-hand-side panels, the ACFs are shown using time-lags covering the full time baseline of 2\,000\,days. Although certain long-term variations can be observed, one cannot visually identify a clear feature associated to the injected magnetic cycle of 1\,000\,days, and so the capability of commonly used kernels to determine features from long-term activity cycles is likely to be low. On the right-hand side, a close-in view of the ACFs at small time-lag is shown to highlight the signature of the stellar rotation. The injected spot lifetime of 100\,days is seen as a general decay of the correlations at longer time-lags. 

\begin{table*}
        \caption{\label{Spodis} Number of spots $N$ created in every 100-day time interval [$t_0,t_0+100$] to mimic a long-term magnetic cycle of 1\,000\,days. We also show the varying mean colatitude of the spots in a northern and in a southern hemisphere for the active longitude spot models, thereby resembling the butterfly diagram.}
        \centering
        \begin{tabular}{ll|cccccccccc}
                \hline \hline \noalign{\smallskip}
                Time interval $t_0$ [d] &  & 0 & 100 & 200 & 300 & 400 & 500 & 600 & 700 & 800 & 900 \\
                $N$ & {\tt ONE}, {\tt TWO}, {\tt RAN}  & 40 & 36 & 32 & 28 & 24 & 20 & 24 & 28 & 32 & 36 \\
                $N$ & {\tt POL}  & 4 & 4 & 3 & 3 & 2 & 2 & 2 & 3 & 3 & 4 \\              
                colat. North [deg] & {\tt ONE}, {\tt TWO} &  125 & 119 & 113 & 107 & 101 & 95 & 101 & 107 & 113 & 119  \\
                colat. South [deg] & {\tt ONE}, {\tt TWO} &  55 & 61 & 67 & 73 & 79 & 85 & 79 & 73 & 67 & 61  \\
                        \noalign{\smallskip} \hline \hline
        \end{tabular}
\end{table*}

For the {\tt ONE} model we see recurring correlations for multiples of $P_{\rm rot}$, with slightly decreasing power over time until the injected spot lifetime of 100\,days. The spotted side of the star, having a width of 60\,deg, smooths out the effect at multiples of $P_{\rm rot}/2$ described by the single-spot models as shown in Fig.\,\ref{ACF1}. This is different from all the other curves. The {\tt POL} distribution is most similar to the single-spot model, although the peaks at $P_{\rm rot}/2$ are not symmetric, there is diminishing power up to $t=100$\,days, and there is essentially no power beyond that time because of the phase shift induced by the new location of the dominant surface half. The same behaviour is seen for the peaks at $P_{\rm rot}$. For the {\tt RAN} distribution, the peaks at $P_{\rm rot}/2$ are higher, because the contrast with the less prominent side of the star is smaller, and they are therefore more strongly correlated and precisely located. The effect is even larger for the {\tt TWO} distribution, where a strong correlation is seen at $P_{\rm rot}/2$. In this case, both peaks diminish in power until $t=100$\,days, but remain present due to the phase stability of the two opposite spotted regions.

In theory, if the two sides of the surface were identical, the kernel peaks would be at the same correlation level and the amplitude of these peaks would be the same. We try to connect this trend with the differences of filling factors of the prominent side ($\mathcal{F}_{\rm prom}$) and its opposite ($\mathcal{F}_{\rm opp}$) and calculate the minimum and maximum values in a constant 25-day window over all the spot maps using the relation of $\mathcal{F}$ and $\kappa_0$ found in Fig.\,\ref{h_calib}. This delivers the mean minimum and maximum filling factors as an approximation of $\mathcal{F}_{\rm opp}$ and $\mathcal{F}_{\rm prom}$, respectively. The numbers are shown in Table\,\ref{ffs}. We note the anti-correlation of $\mathcal{F}_{\rm max} = \mathcal{F}_{\rm prom} - \mathcal{F}_{\rm opp}$ with the relative strength of the peak of the ACFs at $P_{\rm rot}/2$, as seen in Fig.\,\ref{ACF2}. 

\begin{table}
        \caption{\label{ffs} Mean minimum ($\mathcal{F}_{\rm opp}$) and maximum ($\mathcal{F}_{\rm prom}$) filling factors within a 25-day interval for the different spot models. Values are given in \%.}
        \centering
        \begin{tabular}{cccc}
                \hline \hline \noalign{\smallskip}
        Spot configuration & $\mathcal{F}_{\rm opp}$ & $\mathcal{F}_{\rm prom}$ & $\mathcal{F}_{\rm max} = \mathcal{F}_{\rm prom}-\mathcal{F}_{\rm opp}$  \\  \noalign{\smallskip} \hline \noalign{\smallskip}
        {\tt ONE} & 0.0 & 7.1 & 7.5$\pm$2.5 \\
        {\tt POL} & 0.3 & 7.8 & 8.0$\pm$3.8 \\
        {\tt RAN} & 2.8 & 6.3 & 3.5$\pm$1.2 \\
        {\tt TWO} & 2.1 & 4.7 & 2.7$\pm$1.0 \\
                \noalign{\smallskip} \hline \hline
        \end{tabular}
\end{table}

\section{Gaussian process regression} \index{s3} \label{s3}

We apply GP regression to our RV time-series data created for the various spot configurations as introduced in Sect.\,\ref{sec:multispot}, and use the {\tt emcee} code to explore the region of the parameter space with largest likelihoods using a Markov chain Monte Carlo procedure \citep[MCMC,][]{2013PASP..125..306F}. 
The likelihood optimization code needs to formally assign an uncertainty to each measurement. To deal with this, we assign identical \textit{dummy} uncertainties of $\epsilon =10^{-3} \, \sigma_v$ ($\sigma_v$ being the variation in RV), and we force the jitter parameter $\sigma$ (as in Eq.\,\ref{X4}) to be always positive in order to avoid numerical issues (divisions by zero and negative logarithm evaluations). In this sense, the jitter term can be interpreted as the \textit{unexplained variance} that remains because of an imperfect fit, thus providing an additional figure of merit. In the following sections, we review the two most common kernels used in the literature, and describe the interpretation given to their free adjustable parameters. In Sect.\,\ref{s33}, we further present an improved version of the QP kernel, namely the quasi-periodic and cosine (QPC) kernel.

\subsection{Quasi-periodic kernel}  \index{s31} \label{s31}

\begin{figure}
        \centering
        \includegraphics[width=\linewidth]{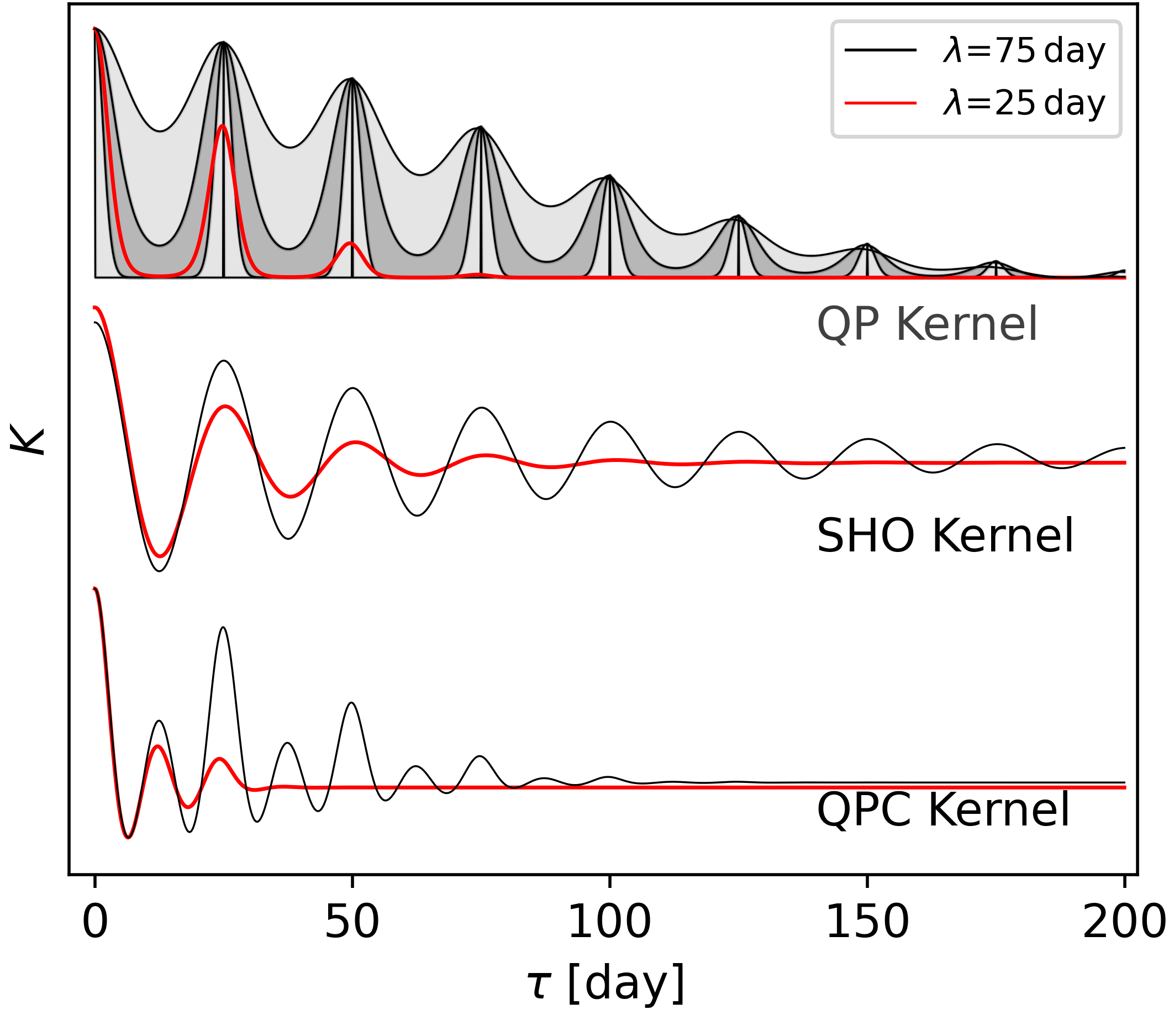}
        \caption{Graphic representation of the three kernels used in this study: the quasi-periodic (QP) kernel as implemented in the {\tt george} code (top), the simple harmonic oscillator (SHO) as implemented in the {\tt celerite} code (middle panel), and the quasi-periodic and cosine (QPC, bottom) kernel introduced in this work, for $P=25$ days. The x-axis is the time-lag $\tau$ between two data points, and the vertical axis is the normalised value of the correlation represented by the kernel. We show the kernels in red for $\lambda=25$\,days ($w=0.31$) and black for $\lambda=75$\,days. For the QP kernel, the light grey area marks $0<w<1$, whereas the dark grey area marks the interval $0.2<w<0.5$, which is more physically motivated. The QPC kernel has $h_2/h_1=1$.}
        \label{Kernels}
\end{figure}

In exoplanet literature, the most commonly used GP kernel to model astrophysical noise is the quasi-periodic (QP) kernel, given as
\begin{equation}
K_{\rm QP}(\tau) =  h^{2} \exp \Big( -\frac{\tau^{2}}{2 \lambda^{2}} - \frac{1}{2{\rm w}^{2}} \sin^{2}(\frac{\pi}{P} \tau) \Big), \label{Ec1}
\end{equation}
\noindent which introduces the four hyper-parameters $h$, $P$, $\lambda$, and $w$. The kernel is implemented in the {\tt george} code by \citet{2015ITPAM..38..252A}. A graphic representation of this kernel with $P=P_{\rm rot}=25$\,days, a range of values for $w$, and two values of $\lambda$ ($75$ and $25$ days) are shown in the upper part of Fig.\,\ref{Kernels}. 

The range of hyper-parameter $w$ is shown in the figure in light grey covering the commonly used prior between 0 and 1. If we inspect Fig.~\ref{Kernels}, we can observe the influence of $w$. Firstly, $w$ reflects the relative strength between the most prominent surface feature and whatever is in the opposite surface half. To match the observed features in the ACF, we examine the simulated ACFs (Fig.\,\ref{ACF2}) in the different configurations. The value of $w$ should be small and of the order of $w<0.5$, otherwise the correlations would not approach zero between peaks as they do in the ACFs of the synthetic data. Furthermore, the value of $w$ influences the width of the peaks of the kernel, which is related to how localised the prominent spots are on the dominant surface half, that is, a small localised spot leads to a narrow peak, periodically repeating at each $P$. If the positions of the most prominent spots change over time or there is an extended group of them, then the width of the peak widens, leading to an approximate lower limit of $w>0.2$. In general, we can translate the hyperparameter $w$ to the physically realistic behaviour of the synthetic ACFs only in this somewhat arbitrary interval $0.2<w<0.5$, as shown by the dark grey area in Fig.\,\ref{Kernels}, including thereby $w \sim 0.5$, which was suggested by \citet{2016AJ....152..204L}.

\begin{figure*}
        \centering
        \includegraphics[width=\linewidth]{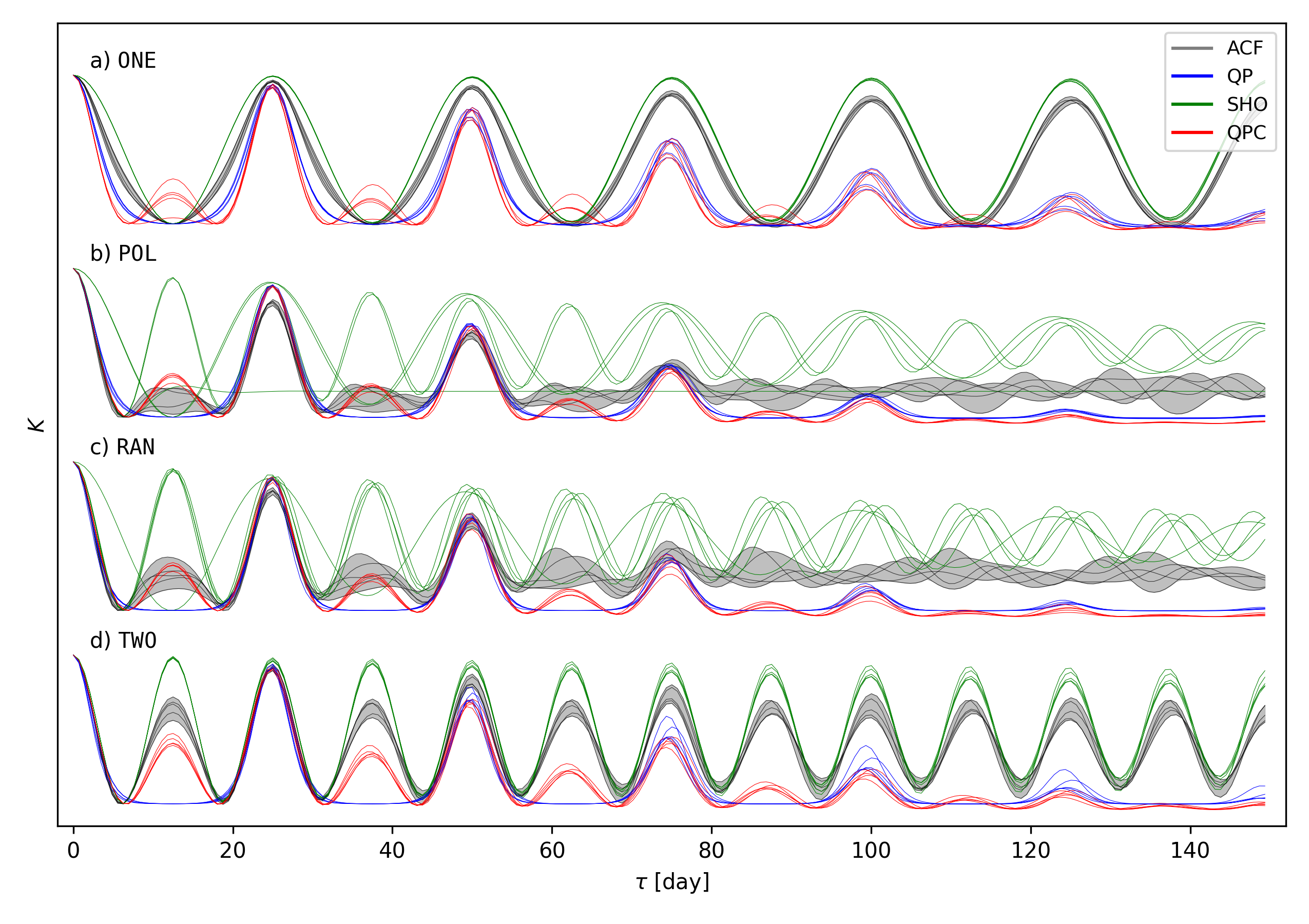}
        \caption{Auto-correlation functions (ACF, grey and black) and GP kernels (QP blue, SHO green, QPC red) with largest fitted likelihoods of all RV data sets sorted by spot configuration (from top to bottom {\tt ONE}, {\tt POL}, {\tt RAN}, and {\tt TWO}). The SHO kernel has a less prominent correlation decay and puts more weight on correlations at larger $\tau$ in comparison to the QP and QPC kernels. We reiterate the fact that the hyperparameters of the kernels are specifically sensitive to the correlations at shorter time-lags because of the linear decrease of the number of data-point pairs $N_{\tau}$ with time-lag $\tau$, i.e. $N_{ \tau}(\tau)=\frac{1-N}{T} \, \tau + N$.}
        \label{ResuP}
\end{figure*}

We maximise the likelihood applying an MCMC procedure with 1\,000 steps on 1\,000 walkers. We set further prior limits to the parameters and hyper-parameters as shown in Table\,\ref{Resu}, where 3.4\,days is the constant temporal distance between two observations, 1/4000\,days is the Nyquist frequency, $\overline{\rm v}$ is the mean RV of the respective data set, and $\sigma_{\rm v}$ its variation. We note in Table\,\ref{Resu} the values leading to the largest likelihood values averaged over the five spot maps for each spot configuration. The hyper-parameters are given alongside with RV offset $\gamma$, the additional RV jitter $\sigma$, and the figures of merit, $\ln{\mathcal L}$ and BIC. In Fig.\,\ref{ResuP}, we compare the best solutions of each kernel with the ACFs of all input RV data sets for the four different spot distributions (from top to bottom) {\tt ONE, POL, RAN}, and {\tt TWO}.

We find that the QP kernel is able to identify the introduced 25-day period for every data set. We also find that there is a clear relation of the hyper-parameter $\lambda$ and the lifetime of spots with $T_{spots} = 100$\,days $= 1.7\pm 0.1 \lambda$. Due to the discrete nature of the problem, the number of data-point pairs $N_{\tau}$ with a certain time-lag $\tau$ is given with $N_{ \tau}(\tau)=\frac{1-N}{T} \, \tau + N$ (with $N$ as the total number of data points, $T$ as the time baseline), which puts more statistical weight on the correlations at short-time-lags. This is why the GP regression always tries to match the most relevant features of the ACF close to $\tau=0$ (which tends to follow a Gaussian $\exp \tau^2/2 \lambda^2$ envelope), thus explaining why the correct value is found despite the poor match of the kernel and the ACF at large time-lags (see right side of Fig.~\ref{ResuP}). The hyper-parameter $\lambda$ is slightly larger for the {\tt ONE} and {\tt TWO} configurations because the spots are always at the same rough longitudes, thus producing coherent correlations on longer time-scales. We find that $w$ is in the range from $0.28$ to $0.36$ for all our experiments with different spot configurations. Moreover, a slight variation of $w$ between those values does not result in significant changes of the likelihoods. As we search for a qualitative translation of the hyperparameters with our test case data, we recommend fixing this parameter to the average $w=0.311\pm0.016$ and removing it as a free parameter of the regression, or at least set priors to [0.2, 0.5], if our aim for the GP modelling is to remove the effect of spots in our RV data. The amplitude $h$ (which shall match the ACF amplitude at zero time-lag $\kappa_{\rm 0, QP}$) is recovered quite consistently, which translates to filling factors $\mathcal{F}_{\rm max} \sim 2\%$. The fitted RV offsets are all consistent with zero. 

In summary, we find that this kernel and its parameters provide a good option for adjusting correlations and physically interpreting them, but we also find that it fails to correctly recover some relevant information (i.e. the $w$ parameter is not sensitive enough to spot configurations). Using the value of the maximum likelihood and the minimum jitter as figures of merit, we find that this kernel performs at its best on configurations {\tt ONE}, {\tt TWO}; and -to a slightly lesser degree- on {\tt RAN} {and \tt POL} configurations. 

\begin{table*}
        \caption{\label{Resu} Largest likelihood solution found for every free adjustable parameter averaged over the five simulations of each spot configuration ({\tt ONE}, {\tt POL}, {\tt RAN}, {\tt TWO}). We show the fitted parameters (from top to bottom) of the QP, SHO, and QPC kernel and calculate filling factors and likelihood differences.}
        \centering
        \begin{tabular}{lcccccc}
                \hline \hline \noalign{\smallskip}
                \hspace{0.3cm} Parameter & {\tt ONE} & {\tt POL} & {\tt RAN} & {\tt TWO} & priors & unit \\ \noalign{\smallskip} \hline \noalign{\smallskip}
                \multicolumn{7}{l}{QP: quasi-periodic kernel} \\ \noalign{\smallskip}
                \hspace{0.3cm} $h=\kappa_{0,QP}$ & 2.94$\pm$0.07 & 3.21$\pm$0.17  & 2.89$\pm$0.10  & 2.89$\pm$0.10 & [0, 5\,$\sigma_v$] & m\,s$^{-1}$ \\
                \hspace{0.3cm} $P$ & 24.96$\pm$0.08 & 24.98$\pm$0.10 & 25.00$\pm$0.09 & 24.92$\pm$0.10 & [3.4, 50]  & day \\
                \hspace{0.3cm} $\lambda$ & 66.98$\pm$4.78 & 51.86$\pm$0.43 & 51.73$\pm$1.34 & 63.06$\pm$5.46 & [3.4, 4000]  & day \\
                \hspace{0.3cm} $w$ & 0.358$\pm$0.011 & 0.320$\pm$0.011 & 0.282$\pm$0.010 & 0.283$\pm$0.013 &  [0,1]  & \\
                \hspace{0.3cm} $\gamma$ & 0.01$\pm$0.16 & $-$0.08$\pm$0.11 & $-$0.02$\pm$0.10 & $-$0.04$\pm$0.03 & [$\overline{v}-$3\,$\sigma_v$, $\overline{v}$+3\,$\sigma_v$] & m\,s$^{-1}$ \\
                \hspace{0.3cm} $\sigma$ & 0.04$\pm$0.08 & 1.37$\pm$0.22 & 0.91$\pm$0.11 & 0.40$\pm$0.07 & [0, 3\,$\sigma_v$]  & m\,s$^{-1}$ \\
                \hspace{0.3cm} $\ln{\mathcal L}$ & $-$1178.0$\pm$33.9 & $-$1507.0$\pm$55.5 & $-$1410.1$\pm$14.4 & $-$1287.0$\pm$8.8 & & \\
                \hspace{0.3cm} BIC & 2394.1$\pm$67.7 & 3052.1$\pm$111.0 & 2858.3$\pm$28.9 & 2612.2$\pm$17.5 & &  \\ \noalign{\smallskip}
                \multicolumn{7}{l}{SHO: simple harmonic oscillator kernel} \\ \noalign{\smallskip}
                \hspace{0.3cm} $C_{0}$ & 14.78$\pm$1.78 & 11.64$\pm$9.42 & 5.47$\pm$0.68 & 10.68$\pm$2.23 & [0, 5\,$\sigma_v$]  & m$^2$\,s$^{-2}$ \\
                \hspace{0.3cm} $P$ & 25.00$\pm$0.02 & 17.34$\pm$6.02 & 14.93$\pm$4.82 & 12.49$\pm$0.01 & [3.4, 50]  & day \\
                \hspace{0.3cm} $P_{\rm life}$ & 1989.4$\pm$455.5 & 89.84$\pm$44.69 & 116.10$\pm$15.27 & 420.8$\pm$148.3 & [3.4, 4000]  & day \\
                \hspace{0.3cm} $\gamma$ & 0.00$\pm$0.00 & 0.00$\pm$0.01 & 0.00$\pm$0.01 & 0.00$\pm$0.01 &  [$\overline{v}-$3\,$\sigma_v$, $\overline{v}$+3\,$\sigma_v$]  & m\,s$^{-1}$ \\
                \hspace{0.3cm} $\sigma$ & 2.72$\pm$0.15 & 3.31$\pm$1.61 & 3.60$\pm$0.21 & 2.91$\pm$0.11 & [0, 3\,$\sigma_v$] & m\,s$^{-1}$ \\
                \hspace{0.3cm} $\ln{\mathcal L}$ & $-$1427.7$\pm$28.7 & $-$1693.2$\pm$55.1 & $-$1604.9$\pm$29.6 & $-$1488.1$\pm$21.4 & & \\
                \hspace{0.3cm} BIC & 2887.1$\pm$57.3 & 3418.1$\pm$110.2 & 3241.5$\pm$59.2 & 3007.9$\pm$42.8 & & \\ \noalign{\smallskip}
                \multicolumn{7}{l}{QPC: quasi-periodic and cosine kernel} \\ \noalign{\smallskip}
                \hspace{0.3cm} $h_1$ & 3.84$\pm$0.11 & 4.40$\pm$0.37 & 3.86$\pm$0.14 & 3.20$\pm$0.15 & [0, 5\,$\sigma_v$]  & m\,s$^{-1}$ \\
                \hspace{0.3cm} $h_2$ & 1.12$\pm$0.51 & 1.97$\pm$0.26 & 1.94$\pm$0.16 & 2.66$\pm$0.26 & [0, 5\,$\sigma_v$]  & m\,s$^{-1}$ \\
                \hspace{0.3cm} $P$ & 24.96$\pm$0.09 & 24.99$\pm$0.09 & 24.99$\pm$0.06 & 24.92$\pm$0.11 & [3.4, 50]  & day \\
                \hspace{0.3cm} $ \lambda$ & 135.32$\pm$9.75 & 102.90$\pm$1.68 & 102.60$\pm$5.74 & 117.25$\pm$4.31 & [3.4, 4000]  & day \\
                \hspace{0.3cm} $\gamma$ & $-$0.03$\pm$0.05 & 0.01$\pm$0.09 & $-$0.03$\pm$0.07 & $-$0.03$\pm$0.07 &  [$\overline{v}-$3\,$\sigma_v$, $\overline{v}$+3\,$\sigma_v$]  & m\,s$^{-1}$ \\
                \hspace{0.3cm} $\sigma$ & 0.04$\pm$0.08 & 1.35$\pm$0.23 & 0.93$\pm$0.11 & 0.40$\pm$0.08 & [0, 3\,$\sigma_v$]  & m\,s$^{-1}$ \\
                \hspace{0.3cm} $\ln{\mathcal L}$ & $-$1179.8$\pm$34.1 & $-$1502.6$\pm$55.5 & $-$1404.2$\pm$15.2 & $-$1266.5$\pm$11.0 & & \\
                \hspace{0.3cm} BIC & 2397.7$\pm$68.3 & 3043.4$\pm$111.0 & 2846.6$\pm$30.4 & 2571.1$\pm$21.9 & & \\ \noalign{\smallskip}
                \multicolumn{7}{l}{Additional parameters} \\ \noalign{\smallskip}
                \hspace{0.3cm} $\mathcal{F}_{\rm max, QP}$ & 1.8$\pm$0.1 & 1.9$\pm$0.1 & 1.8$\pm$0.1 & 1.8$\pm$0.1 & & \% \\
                \hspace{0.3cm} $\mathcal{F}_{\rm max, SHO}$ & 2.3$\pm$0.1 & 2.1$\pm$0.8 & 1.4$\pm$0.1 & 2.0$\pm$0.2 & & \% \\
                \hspace{0.3cm} $\mathcal{F}_{\rm max, QPC}$ & 2.4$\pm$0.1 & 2.9$\pm$0.2 & 2.6$\pm$0.1 & 2.5$\pm$0.1 & & \% \\
                \hspace{0.3cm} $\mathcal{F}_{\rm opp}/\mathcal{F}_{\rm prom}$ & 28.1$\pm$11.7 & 40.8$\pm$5.4 & 45.0$\pm$3.2 & 63.9$\pm$4.1 & & \% \\  \noalign{\smallskip}
                \hspace{0.3cm} $\sqrt{C_{0}}=\kappa_{\rm 0,SHO}$ & 3.84$\pm$0.23 & 3.41$\pm$1.38 & 2.34$\pm$0.15 & 3.27$\pm$0.34 &    & m\,s$^{-1}$ \\
                \hspace{0.3cm} ($h_1^2+h_2^2)^{\frac{1}{2}} =\kappa_{\rm 0,QPC} $ & 4.00$\pm$0.18 & 4.82$\pm$0.35 & 4.32$\pm$0.14 & 4.17$\pm$0.20 & & m\,s$^{-1}$ \\   \noalign{\smallskip}
                \hspace{0.3cm} $\Delta \ln{\mathcal L}_{\rm QP-SHO}$ & $-$250 & $-$186 & $-$195 & $-$201 & & \\
                \hspace{0.3cm} $\Delta \ln{\mathcal L}_{\rm QP-QPC}$ & $-$1.8$\pm$2.2 & 4.4$\pm$1.7 & 5.9$\pm$2.6 & 20.6$\pm$4.5 & & \\
                \hspace{0.3cm} $\Delta \ln{\mathcal L}_{\rm QP-QPCw}$ & 1.5$\pm$2.1 & 4.6$\pm$1.8 & 7.3$\pm$3.0 & 22.6$\pm$4.8 & & \\  \noalign{\smallskip}
                \hspace{0.3cm} $\Delta {\rm BIC}_{\rm QP-SHO}$ & 493 & 366 & 383 & 396 & & \\
                \hspace{0.3cm} $\Delta {\rm BIC}_{\rm QP-QPC}$ & 3.6$\pm$4.3 & $-$8.7$\pm$3.4 & $-$11.7$\pm$5.3 & $-$41.1$\pm$9.0 & & \\
                \hspace{0.3cm} $\Delta {\rm BIC}_{\rm QP-QPCw}$ & 3.3$\pm$3.8 & $-$2.9$\pm$3.0 & $-$8.1$\pm$4.6 & $-$38.8$\pm$8.2 & & \\
                \noalign{\smallskip} \hline \hline
        \end{tabular}
\end{table*}

\subsection{Simple harmonic oscillator kernel}   \index{s32} \label{s32}

We also study the stochastically driven simple harmonic oscillator (SHO) kernel implemented in the {\tt celerite} algorithm by \citet{2017AJ....154..220F} and written as
\begin{eqnarray}
K_{\rm SHO}(\tau) &=& C_{0} e^{-\tau /P_{\rm life}} \Big[ \cos{(\frac{q}{P}  \tau)} +  \frac{P}{q P_{\rm life}} \sin{(\frac{q}{P} \tau )} \Big],\\ \label{Ec2}
q &=& 2 \pi \sqrt{\Big(2 \pi \frac{P_{\rm life}}{P} \Big)^{2}-1},
\end{eqnarray}
\noindent with $P<2\, \pi \,P_{\rm life}$. This kernel introduces one less hyper-parameter than the QP kernel, with $C_0$, $P$, and $P_{\rm life}$, whereas the  parameters that are fitted in the case of the {\tt celerite} code are the logarithms of
\begin{equation}
S_0 = \frac{C_0}{2 P_{\rm life}} \Big(\frac{P}{\pi}\Big)^2 ; \, \, \, \, w_0 = \frac{2 \pi}{P} ;  \, \, \, \, Q = \pi \frac{P_{\rm life}}{P}. \label{Ex2}
\end{equation}
\noindent An illustration of the kernel is given in the middle part of Fig.\,\ref{Kernels} using hyper-parameter $P=25$\,days, and $P_{\rm life}=25$ (red) and 75\,days (black). Here, $P$ results in the sinusoidal variation while $P_{\rm life}$ is responsible for a slower exponential decay, and hence a stronger correlation for larger time-lags $\tau$ compared to the QP kernel. However, the most significant difference is the presence of negative values in this kernel, resulting in anti-correlations for opposite surface halves over time. $C_0$ is given as $K_{\rm SHO}(\tau = 0) = \kappa^2_{\rm 0, SHO}$ in m$^2$\,s$^{-2}$.

We apply the same MCMC approach as before and show the results of the GP regression using the SHO kernel in Table\,\ref{Resu} and Fig.\,\ref{ResuP}. The filling factors connected to the hyper-parameter $C_0$ are for all the different spot configurations slightly larger than for the QP kernel. The more consistent solutions for the five spot maps are given with the {\tt ONE} spot configuration, where a $P=P_{\rm rot}=25$\,day period is found, and {\tt TWO}, where $P=P_{\rm rot}/2$ is visible. In those two cases, where the spots are always distributed around the same longitudes and the $P_{\rm rot}$-periodicity in the ACFs is stable over time (black curves), the hyper-parameter $P_{\rm life}$ is rather large and unconstrained. For the {\tt POL} and {\tt RAN} configurations, the SHO kernel and its $P$ hyper-parameter is not able to consistently distinguish between $P_{\rm rot}$ and $P_{\rm rot}/2$. However, as the periodicity is lost after a certain time for both configurations, the hyper-parameter $P_{\rm life}$ is identified consistently with $\sim$100\,days. For the {\tt POL} case there is even a solution with $P_{\rm life}=8\,$days. $P_{\rm life} < P$ is a behaviour which we often find in our studies and in the literature of unevenly sampled RV data. When this happens, the value of $P$ is no longer representative of the real $P_{\rm rot}$ and a kernel with only an exponential or a Gaussian profile should be used instead to model correlations. 

Large additional jitters $\sigma$ of up to 3.6\,m\,s$^{-1}$ are needed to fit the data and, despite having one less hyper-parameter than the QP kernel, BIC values are much larger. Given all these interpretation problems, and the fact that the fits are generally poorer (in terms of both lower likelihoods and higher jitter), we advise against using SHO kernels when studying correlations caused by stellar activity in RV time-series. 

\subsection{Quasi-periodic and cosine  kernel} \index{s33} \label{s33}

Given the shape of the ACFs found in our simulations, we propose a new kernel that we implemented to run in the {\tt george} code (.yml file given in appendix\,\ref{aa}), namely a combination of the QP kernel and a damped cosine function, dubbed a quasi-periodic and cosine (QPC) kernel, with,
\begin{eqnarray}
& & K_{\rm QPC}(\tau)  =  \exp \Big( -2\frac{ \tau^{2}}{\lambda^{2}} \Big) \\
& & \cdot \Bigg[    h_1^2 \, \exp \Big(- \frac{1}{2 \, w_0^2} \, \sin^{2}{(\pi \frac{\tau}{P})} \Big) + h_2^2 \, \cos{(4\pi \frac{\tau}{P})}  \Bigg]. \nonumber\label{Ec3}
\end{eqnarray}
\noindent This kernel introduces four hyper-parameters, $P$, $h_1$, $h_2$, and $\lambda$. The periodic peaks of the QPC kernel are again given with $P$, whereas the cosine function adds a periodicity at $P/2$. Here, $h_1$ and $h_2$ are the amplitudes of the QP component and of the cosine function, respectively, with $\kappa^2_{\rm 0, QPC}=h_1^2+h_2^2$. The kernel is shown in Fig.\,\ref{Kernels} with $P=25$\,days, $h_1/h_2=1$, and $\lambda=25$ (red), and 75\,days (black). This kernel takes advantage of the functional dependence at small time-lags $\sim \exp(- \tau^2 / \lambda^2)$ found in the QP kernel (Eq.\,\ref{Ec1}), and includes a cosine to account for the feature typically appearing at P/2. We adapted $\lambda_{\rm QPC}= 2 \lambda_{\rm QP}$ so that this hyper-parameter describing the exponential decay of the correlation represents the average spot lifetime more closely. Given its almost null impact in the quality of the fits, we further consider $w$ as a fixed parameter with $w=w_0=0.31$. Nevertheless, we run our models also with free $w$, as a fifth GP hyper-parameter, and show the statistics in Table\,\ref{Resu} (QPCw).

In Table\,\ref{Resu} and Fig.\,\ref{ResuP}, we observe that both hyper-parameters $P$ and $\lambda$ consistently match the introduced $P_{\rm rot}=25$\,days and spot lifetime $T_{\rm spot}=100\,$days. Offsets are consistent with zero. Additional jitters are small and behave similarly to the values found for the QP kernel. Amplitudes and filling factors are instead the largest of the three kernels, reaching almost 3\% spot coverage. Whereas $h_1$ is relatively consistent, we see a very different behaviour for the amplitude $h_2$, which is directly connected to the contribution of the cosine at $\tau=P_{\rm rot}/2$. The relationship between the filling factors of the prominent stellar surface half ($\mathcal{F}_{\rm prom}$), which is calculated by both amplitudes with $(h_1^2+h_2^2)^\frac{1}{2} = \kappa_{\rm 0, QPC}$, and the opposite half ($\mathcal{F}_{\rm opp}$) with $h_2$ strongly depends on the spot distribution model. It yields $\mathcal{F}_{\rm opp}/ \mathcal{F}_{\rm prom}$=28, 41, 45, and 64\,\% for  {\tt ONE, POL, RAN}, and {\tt TWO} configurations, respectively, accounting for the increasing correlation of the peak at $P_{\rm rot}/2$. We find a decrease in BIC compared to the QP kernel with increasing $h_2$ and significantly better models (dBIC$<-10$). The BIC values are larger for an additional parameter $w$ (QPCw case), which again indicates that these parameters shall be set to a constant value.

\section{Conclusions} \index{s4} \label{s4}

In this paper we begin by deriving the basic relations between a generating signal, the covariances it induces, and their relation to Gaussian process kernels. We then calculate the auto-correlation functions of two simple analytic cases to illustrate the physical meaning of the parameters in the resulting kernels. With our {\tt StarSim} code, we create synthetic RV time-series data for a rotating test-case M-dwarf star with a central single spot, and with four different types of evolving spot maps (which we call configurations) including: one active longitude, two active longitudes, a random spot distribution, and a large polar spot distribution. We use simulated data to explore the relation of the ACFs with stellar inclination and spot filling factor. We then apply GP regression with the commonly used QP and SHO kernels in order to calibrate the physical meaning of the hyper-parameters often presented in the literature. Because of the different imprints of the spot configurations on the ACFs and as a result of this study, we propose the use of a new quasi-periodic and cosine (QPC) kernel. Our main results can be summarised as follows:

\begin{itemize}

    \item \textbf{Generating signals of any nature produce covariances} in time-series. These can be computed from models and/or observations using the definition of the ACF. This leads to a physical interpretation of the various adjustable kernel hyper-parameters often used in the literature.
    
    \item \textbf{The ACF of single-spot configurations} shows its main peaks at multiples of $P_{\rm rot}$ and is close to zero at multiples of $P_{\rm rot}/2$. It shows further characteristic lobes of negative correlations, which vary with different characteristics of the star and spots, and which are not accounted for in the commonly used GP kernels. Its amplitude $\kappa_0$ depends linearly on the filling factor $\mathcal{F}_{\rm max}$ and quadratically on the stellar inclination angle $\sin i$.
    
    \item \textbf{The auto-correlation functions of multi-spot configurations} explicitly show that correlations generally diminish over a certain timescale connected to the spot lifetime. The repeatability of peaks in the ACF at $P_{\rm rot}$ and $P_{\rm rot}/2$ is connected to the phase stability of the spot configurations, that is, it remains very strong over very long timescales if the spot longitude remains roughly the same. Peaks in the ACF at $P_{\rm rot}/2$ are always found in {\tt ONE, POL, RAN}, and {\tt TWO} spot configurations, where the contrast of these peaks to the correlation at $\tau=0$ is connected to the filling factor difference between the two halves of the stellar surface.
    
    \item \textbf{Quasi periodic (QP) kernel} GP regression accounts for the ACF peaks at $P_{\rm rot}$, but not the ones at $P_{\rm rot}/2$. The decay hyper-parameter $\lambda$ and the lifetime of the simulated spots are clearly mapped to each other. We argue that, despite the fact that the kernel generally does a poor job adjusting the shape of the ACF at large time-lags, it still identifies the correct lifetime of the spots because of the larger number of pairs of data points at shorter time-lag contributing more to the likelihood function. As expected, we find that $\kappa_0$ from the simulations is mapped one-to-one to the kernel hyper-parameter $h$. The parameter $w$ behaves as a form factor and should always have  values in the range $0.2<w<0.5$; its introduction as a free parameter does not have a physical translation, nor does it improve the likelihood, and so we recommend fixing it at a reference value of $w_0 = 0.31$ when modelling spot-induced activity with GPs in RV data.
    
    \item \textbf{Simple harmonic oscillator (SHO) kernel} GP regression produces unreliable estimates of the rotation periods, typically obtaining values between $P_{\rm rot}$ and $P_{\rm rot}/2,$ strongly depending on the spot configurations. The amplitude $C_0$ matches the value of the ACF $\kappa_0^2$ at zero time-lag. However, we find that the decay parameter $P_{\rm life}$ in the SHO varies between large values and the introduced spot lifetime, and therefore its value is not a reliable measure of the typical lifetimes of stellar activity features. Overall, the kernel does not consistently recover the introduced parameters or the shape of the ACF at small time-lags, and produces fits of substantially poorer overall quality. This is reflected in lower maximum-likelihood values and the need for higher jitter values when compared with the other kernels discussed here.  
    
    \item \textbf{Our proposed quasi-periodic and cosine (QPC) kernel} is designed to use the best features of the QP kernel and adds an additional term to account for the $P_{\rm rot}/2$ peaks in the ACFs. It has the same number of parameters as the QP kernel, and regression experiments find $\kappa^2_{\rm 0,QPC} = h_1^2+h_2^2 $. The value of $h_2$ is related to the strength of the ACF peak at $P_{\rm rot}/2$, and and can therefore be used to distinguish between different spot configurations. As for the QP kernel, the periodicity of $P_{\rm rot}$ and the spot lifetime ($\lambda_{\rm QPC}= 2 \lambda_{\rm QP}$) are found to be consistent with the corresponding simulated quantities. The better performance of the QPC kernel is demonstrated by the fact that it leads to significantly higher likelihoods (and smaller jitters) for most of our test cases.

\end{itemize}

We studied the qualitative connection between the auto-correlation functions of different spot configurations and the hyperparameters of different Gaussian process kernels. For this exercise, we used evenly sampled errorless RV data of a spotted rotating edge-on M-dwarf, with no differential rotation and no faculae, and fixed parameters such as the spot lifetime or temperature differences. We did not calculate Bayesian evidence in the form of marginal likelihoods, but relied rather on the qualitative behaviour of the synthetic curves and on the best fit of a Markov chain of Gaussian processes, including the best data likelihoods, additional jitters, and the controversial BIC, averaged over five iterations of stellar spot maps for each spot distribution. If we were to apply the exercise to data with for example realistic measurement uncertainties, uneven sampling, and so on, better statistics could certainly be achieved in some cases by including additional free parameters such as for example the form factor $w$, but we would lose the physical translation with the evolving dark spots on the rotating stellar surface as obtained by this qualitative study.

We plan to further extend our studies from the RVs discussed here to other observable quantities (line widths, chromospheric emission, photometry, etc.), and their corresponding simulated output using {\tt StarSim}. In addition to this calibration of hyper-parameters, it should be possible to produce improved kernels and to extract more physical information from combined measurements. We also plan to amplify our study to quantify the imprint on the RV data of different stellar parameters such as stellar type, where we expect the same qualitative results. The more closely the analysis of stellar activity can be related to physical processes and quantities, the more able we are to use observational and analysis techniques to mitigate (and eventually `model-out') the \textit{noise} of astrophysical origin in high-precision exoplanet searches.

\begin{acknowledgements}
The authors acknowledge support from the Spanish Ministry of Science and Innovation and the European Regional Development Fund through grant PGC2018-098153-B- C33, as well as the support of the Generalitat de Catalunya/CERCA programme. 
\end{acknowledgements}

\bibliography{bibtex}{}
\bibliographystyle{aa}

\newpage
\onecolumn

\begin{appendix}

\section{Quasi-periodic and cosine kernel C++ code as implemented in the {\tt george}  algorithm} \index{aa} \label{aa}

name: ExpSine2CosineKernel \\
stationary: false \\
params: [per, h1, h2, lam] \\
\newline        
reparams:\\
\hspace*{0.5cm} f1: return -2.0 /lam/lam;\\
\hspace*{0.5cm} f2: return M\_PI/per;\\
\hspace*{0.5cm} ww: return 0.31*0.31;\\
\newline       
value: |\\
\hspace*{0.5cm} double factor1 = exp(f1 *(x1-x2)*(x1-x2)); \\
\hspace*{0.5cm} double factor2 = h1*h1 * exp( -sin(f2*(x1-x2)) * sin(f2*(x1-x2)) /2.0/ww );\\
\hspace*{0.5cm} double factor3 = h2*h2 * cos(4.0*f2*(x1-x2));\\
\hspace*{0.5cm} return factor1*(factor2+factor3);\\
\newline
grad:\\
\hspace*{0.5cm} per: |\\
\hspace*{1cm} double factor1 = exp(f1 *(x1-x2)*(x1-x2));\\
\hspace*{1cm} double factor2 = h1*h1 * exp( -sin(f2*(x1-x2)) * sin(f2*(x1-x2)) /2.0/ww );\\
\hspace*{1cm} double N1 = sin(f2*(x1-x2));\\
\hspace*{1cm} double N2 = cos(f2*(x1-x2));\\
\hspace*{1cm} double N3 = h2*h2*sin(4.0*f2*(x1-x2));\\
\hspace*{1cm} return factor1*f2*(x1-x2)/per*(factor2/ww*N1*N2 + 4.0*N3);\\     
\hspace*{0.5cm} h1: |\\
\hspace*{1cm} double factor1 = exp(f1 *(x1-x2)*(x1-x2));\\
\hspace*{1cm} double factor2 = h1*h1 * exp( -sin(f2*(x1-x2)) * sin(f2*(x1-x2)) /2.0/ww );\\
\hspace*{1cm} return 2.0/h1*factor1*factor2;\\        
\hspace*{0.5cm} h2: |\\
\hspace*{1cm} double factor1 = exp(f1 *(x1-x2)*(x1-x2));\\
\hspace*{1cm} double factor3 = h2*h2 * cos(4.0*f2*(x1-x2));\\
\hspace*{1cm} return 2.0/h2*factor1*factor3;\\  
\hspace*{0.5cm} lam: |\\
\hspace*{1cm} double factor1 = exp(f1 *(x1-x2)*(x1-x2));\\
\hspace*{1cm} double factor2 = h1*h1 * exp( -sin(f2*(x1-x2)) * sin(f2*(x1-x2)) /2.0/ww );\\ 
\hspace*{1cm} double factor3 = h2*h2 * cos(4.0*f2*(x1-x2)); \\ 
\hspace*{1cm} return -2.0*f1/lam * (x1-x2)*(x1-x2)*factor1*(factor2+factor3);\\                  
\hspace*{0.5cm} x1: |\\
\hspace*{1cm} double factor1 = exp(f1 *(x1-x2)*(x1-x2));\\
\hspace*{1cm} double factor2 = h1*h1 * exp( -sin(f2*(x1-x2)) * sin(f2*(x1-x2)) /2.0/ww );\\
\hspace*{1cm} double factor3 = h2*h2 * cos(4.0*f2*(x1-x2)); \\ 
\hspace*{1cm} double N1 = sin(f2*(x1-x2));\\
\hspace*{1cm} double N2 = cos(f2*(x1-x2));\\
\hspace*{1cm} double N3 = h2*h2*sin(4.0*f2*(x1-x2));\\  
\hspace*{1cm} return -factor1* ( 2.0*(x1-x2)*f1*(factor2+factor3) + f2*h1*h1*x1/ww*N1*N2- 4.0*x1*N3);\\
\hspace*{0.5cm} x2: |\\
\hspace*{1cm} double factor1 = exp(f1 *(x1-x2)*(x1-x2));\\
\hspace*{1cm} double factor2 = h1*h1 * exp( -sin(f2*(x1-x2)) * sin(f2*(x1-x2)) /2.0/ww );\\
\hspace*{1cm} double factor3 = h2*h2 * cos(4.0*f2*(x1-x2)); \\ 
\hspace*{1cm} double N1 = sin(f2*(x1-x2));\\
\hspace*{1cm} double N2 = cos(f2*(x1-x2));\\
\hspace*{1cm} double N3 = h2*h2*sin(4.0*f2*(x1-x2));\\
\hspace*{1cm} return factor1* ( 2.0*(x1-x2)*f1*(factor2+factor3) + f2*h1*h1*x1/ww*N1*N2- 4.0*x1*N3); \\

\end{appendix}

\end{document}